\documentclass[apj,numberedappendix]{emulateapj}

\usepackage{apjfonts}

\usepackage{graphicx}
\usepackage{color}
\usepackage{amsbsy}
\usepackage{mathrsfs}
\usepackage{mathpazo,bm}
\usepackage{amsmath}

\newlength{\hfwidth}
\newlength{\hfwidthsingle}
\newlength{\figspace}
\newcommand{\beq}{\begin{equation}}
\newcommand{\eeq}{\end{equation}}

\renewcommand{\v}[1]{{\boldsymbol{#1}}} 

\newcommand{\del}{\v{\nabla}}

\newcommand{\Div}{\del\cdot}

\newcommand{\degree}{\ensuremath{^\circ}}
\newcommand{\degreep}{\ensuremath{^\circ}~}

\newcommand{\Fig}[1]{Fig.~\ref{#1}}
\newcommand{\fig}[1]{\Fig{#1}}

\newcommand{\sect}[1]{Sect.~\ref{#1}}

\defcitealias{Richert+15}{Paper I}
\defcitealias{Lyra+16}{Paper II}

\definecolor{brown}{rgb}{0.42,0.24,0.07}
\definecolor{darkgreen}{rgb}{0.0,0.6,0.00}
\definecolor{purple}{rgb}{0.7,0.0,0.7}
\definecolor{black}{rgb}{0.0,0.0,0.0}

\def\black#1{\textcolor{black}{#1}}

\newcommand{\mm}[1]{\black{ #1}}
\newcommand{\ed}[1]{\black{ #1}}

\shorttitle{Observational Signatures}
\shortauthors{Hord et al.}

\slugcomment{Draft version}

\begin{document}

\title{On shocks driven by
  high-mass planets in radiatively inefficient
  disks. III. Observational signatures in thermal emission and scattered light.}
\author{Blake Hord\altaffilmark{1,2,3}, Wladimir Lyra\altaffilmark{2,3}, \\
Mario Flock\altaffilmark{3}, Neal J. Turner\altaffilmark{3}\ed{, and Mordecai-Mark Mac Low\altaffilmark{4}}}
\altaffiltext{1}{Dobbs Ferry High School, 505 Broadway, Dobbs Ferry,
  NY 10522, \ed{blake.hord@dfsd.org}}
\altaffiltext{2}{California State University, Northridge. Department of Physics and Astronomy 18111 Nordhoff St, Northridge, CA 91330.}
\altaffiltext{3}{Jet Propulsion Laboratory, California Institute of Technology, 4800
Oak Grove Drive, Pasadena, CA, 91109.}
\altaffiltext{\ed{4}}{\ed{Department of Astrophysics, American Museum of Natural History, Central Park West at 79th Street, New York, NY 10024-5192.}}
\date{Received ; Accepted}

\begin{abstract}
Recent observations of the protoplanetary disk around the Herbig Be
star HD 100546 show two bright features in infrared ($H$ and
$L^\prime$ bands) at about 50 AU\ed{, with one so far unexplained.} 
We explore the observational signatures of a high mass planet causing shock
  heating 
in order to determine if it could be the
  source of the 
\mm{unexplained infrared feature} in HD 100546. More fundamentally, we
identify and characterize planetary shocks as an extra, hitherto ignored, source of luminosity in transition disks. 
The RADMC-3D code is used to perform dust radiative transfer
  calculations on the hydrodynamical disk models, \mm{including
    volumetric heating}. 
A stronger shock heating rate by a factor 20 
\mm{would be} 
   necessary \ed{to \mm{qualitatively} reproduce the
   morphology of the second infrared source}. Instead, we find
  that the outer 
\mm{edge of the gap carved by the planet heats up} by about 50\% relative to the
  initial reference temperature, which leads to an increase in the
  scale height. 
The \mm{bulge is illuminated by the central star, producing a lopsided
  feature in scattered light}, as the outer gap edge shows an asymmetry in density and 
  temperature \ed{attributable to} a secondary spiral arm launched
  not from the Lindblad resonances but from the 2:1 \ed{resonance}. 
We conclude that high-mass planets 
  lead to shocks in disks that may be \mm{directly} observed,
  \mm{particularly at wavelengths of 10~$\mu$m or longer,}
  \mm{but that they are more likely to reveal their}
  presence in scattered light by puffing up their outer gap edges and exciting multiple spiral arms.
\end{abstract}


\section{Introduction}
\label{sect:introduction}

Decades of analytical investigation and numerical simulations of
planets in circumstellar disks have shown that
planet-disk interaction leads to the development of one-armed spirals
\citep{GoldreichTremaine79,LinPapaloizou79,LinPapaloizou93,Kley1999,PapaloizouNelson05,deValBorro+06,Baruteau+14}, which drive angular momentum transport and planet
migration. As the resolution of recent observations increase and we
get a clearer view of the processes happening during planet formation
in circumstellar disks, spiral structures feature as one of the most
common and prominent results of these observations \citep{Muto+12,Garufi+13,Benisty+15}. As a spiral is a common
hydrodynamical response of planet-disk interaction, this exciting 
hypothesis was naturally raised as a likely interpretation of the
observations. However, more concrete evidence is required 
to solidify that possibly coincidental connection. 	


The circumstellar disk around the star HD 100546,
\mm{shows spiral-like features}
\citep{Currie+14} with photometric temperature of 465 $\pm$ 40K
\citep{Lyra+16} given the magnitudes at the $L^\prime$
\citep{Currie+14} and $H$ \citep{Currie+15} wavebands\ed{, with measurements indicating weak polarization and thus thermal emission.} 


\citet[hereafter Paper I]{Richert+15} simulated the wakes of high-mass planets in non-isothermal
2D disks, finding that unless cooling is efficient, the shocks
introduce enough entropy in the flow that the wake becomes
unstable to buoyancy and the disk develops turbulence. 
Effectively, the gravitational potential well of the planet powers a vigorous heat source.



\citet[hereafter Paper II]{Lyra+16} extended the calculation to three dimensions,
simulating a 5M$_J$ planet embedded in a non-isothermal disk. 
\mm{In a three-dimensional disk, gas hit by a shock propagating
  parallel to the disk midplane can
  expand vertically, forming a shock bore}
\ed{\citep{GomezCox04,BoleyDurisen06}}. 
After losing significant energy in the cooler upper atmosphere
of the disk, the gas descends back into the midplane again leading to
a turbulent surf around the planet orbit. In the midplane, two hot
shock lobes form in which the temperature rises to $\approx$500\,K,
generally matching the temperature needed to explain the infrared emission in HD
100546, if thermal. 

However, 
\mm{this is not a complete explanation.} While the photometric temperature in
the source around HD 100546 is very similar to the temperature we
find in our model for the planet's hot Lindblad lobes, \ed{in HD 100546
the emission emanates from the disk surface, whereas in our model} it lies in the
midplane. In \citetalias{Lyra+16} we could not reasonably reproduce the
temperature in the upper atmosphere of the disk because we used a 
simple approach to the energy equation, namely Newton cooling with a
cooling rate parametrized as a function of the optical depth. Instead
of warming up the layers above, the heat emitted by the shocks simply
disappears as it reaches the upper, optically thin, atmosphere. In
reality that radiation would have been reabsorbed by other parts of
the disk, including the disk surface where the observed radiation
originates. 

In this work, we bridge this gap by performing full radiative transfer
post-processing with RADMC-3D \citep{Dullemond+12}, a popular Monte Carlo radiative
transfer software. We compute images based on the simulations
presented in \citetalias{Lyra+16},
and compare them to the observations. This is also the first work to include shock
heating as a source of energy in simulated observations of
protoplanetary disks. This paper is organized as follows. In
\sect{sect:methods} we 
\mm{describe} our methods, in \sect{sect:results}
we present the results, in \sect{sect:discussion} a discussion, followed by a conclusion in
\sect{sect:conclusions}.

\section{Methods}
\label{sect:methods}

We use the radiative transfer code RADMC-3D \citep{Dullemond+12} to determine the
temperature in the disk. The RADMC-3D code 
 \mm{combines} a Monte Carlo code based
on the work of \cite{BjorkmanWood01}, 
\mm{with a ray-tracing mode to simulate observations of the resulting
  temperature distribution.} This 
\mm{technique}
provides a more realistic treatment of cooling than 
the hydrodynamical calculation of \citetalias{Lyra+16}
discussed above, 
and \mm{in particular} works well in determining the temperature of
the photosphere.
\mm{In practice, we}
first use thermal Monte Carlo simulations to determine the temperature of the optically thin regions of the disk. Then, a 
ray-tracing computation is used
to create a synthetic image of what would be 
observed at infrared wavelengths \mm{of interest}.

The hydrodynamical model was calculated with the Pencil Code\citep{BrandenburgDobler02,BrandenburgDobler10}. All required input for the RADMC-3D thermal Monte Carlo simulation was converted from the 
Pencil data through a newly created pipeline between the two codes. The last snapshot (at 41 orbits
of the planet) was taken from the Pencil Code simulation and its parameters were input into 
RADMC-3D.

RADMC-3D requires the following inputs to compute the thermal Monte Carlo simulation: the 
grid size, dust density, wavelength-dependent dust opacity, heating rate, star size, and star location. 

The grid was taken directly from the Pencil Code, creating a spherical grid with dimensions 
(\textit{$N_r$,$N_\theta$,$N_\phi$})=(256, 128, 768). The radial
dimension spans the range 
$[0.4,2.5]$. With the code \mm{length} unit being 5 AU, this is equivalent to [2.1,13]~AU. The meridional 
dimension ranges from $[-0.28,0.28]$ radians, equivalent to $4H$
above and below the midplane, where $H=\varOmega_k/c_s$ is the pressure scale height,
$\varOmega_k$ is the Keplerian angular frequency and $c_s$ the sound speed. The azimuthal 
dimension ranges from 0 to 2$\pi$.

We assume well-coupled 
\mm{micrometer-sized} 
dust grains, perfectly coupled to
the gas. The dust density then follows directly from the Pencil Code
output, scaled down by a factor 100 (the dust-to-gas ratio) and
converted into 
\mm{cgs units of grams per cubic centimeter.}

\begin{figure}
  \begin{center}
    \resizebox{\columnwidth}{!}{\includegraphics{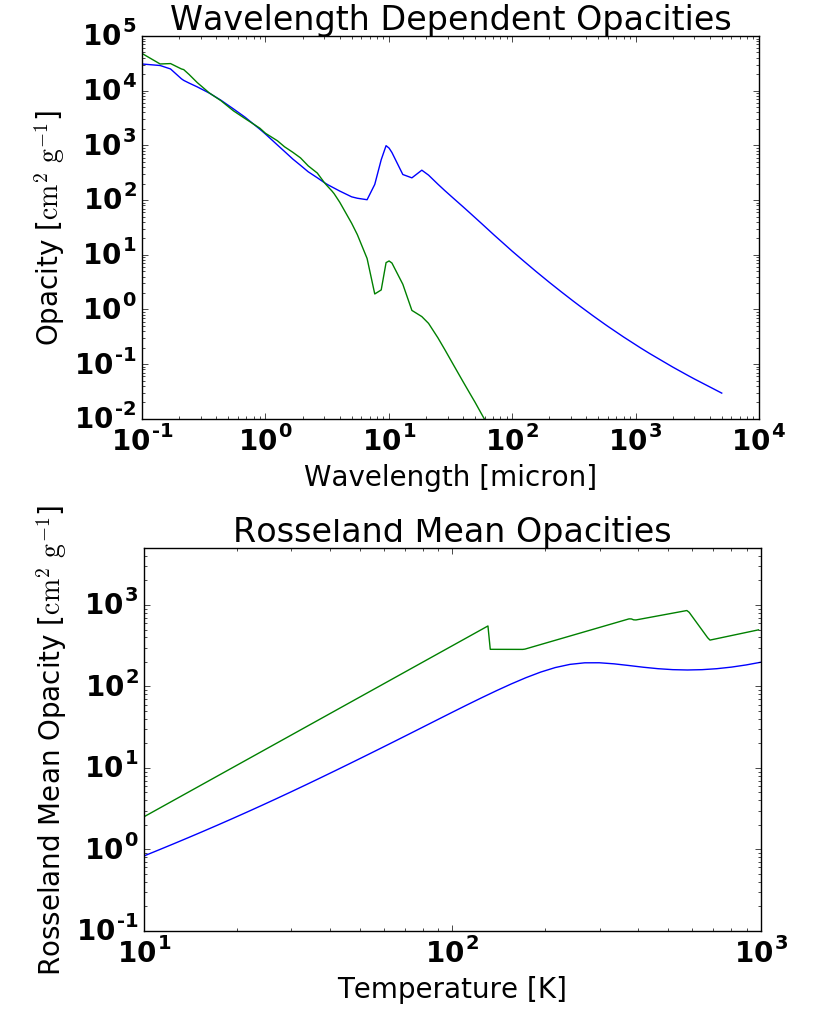}}
\end{center}
\caption[]{The wavelength-dependent opacities from
  \cite{Preibisch+93}, including the absorption (top, blue) and 
scattering (top, green) opacities, input into RADMC-3D. The calculated Rosseland mean opacities (bottom, blue) match the 
Rosseland mean of \cite{Bell97}. The piece-wise Rosseland
Mean opacity based on \cite{Dangelo+03} and 
implemented in the Pencil Code (bottom, green) only varies by at most a factor two from the calculation.}
\label{fig:input-opacity}
\end{figure}

The wavelength-dependent opacity is that of \cite{Preibisch+93}. We use the opacities for the temperature regime from 
125\,K, the water-ice sublimation threshold, to 1500\,K, the silicate sublimation threshold, shown on the top of
\fig{fig:input-opacity}. The calculated Rosseland mean opacity
(\fig{fig:input-opacity}, bottom) closely matches the piece-wise 
Rossland mean opacity from \cite{Bell97}.

One of the novel aspects of our work is that, in addition to stellar illumination, the shock heating rate is used
as extra heating rate input. This heating rate is given by 

\begin{equation}
  \mathcal{H}_{\rm sh} = \rho \nu_{\rm sh} \left(\Div\v{u}\right)^2.
\label{eq:shocktemp}
\end{equation}

\noindent where $\nu_{\rm sh}$ is an artificial shock viscosity,
proportional to the positive part of $\Div{\v{u}}$ according to
 
\begin{equation}
  \nu_{\rm sh} = c_{\rm sh} \left<\max_3[(-\Div\v{u})^+]\right>{\left[\min(\Delta x)\right]}^2{.}
  \label{eq:shock-visc}
\end{equation}

Our treatment of shocks has been
   described in \cite{Haugen+04} and detailed in \citetalias{Richert+15} and \citetalias{Lyra+16}. In Paper I we experimented with a range of values for
the coefficient $c_{\rm sh}$, and found that it did not affect
the result of the simulation as long as
it is sufficient to smooth the shock into a resolvable length for the 
stencil (5 zones in each direction). When the shock is resolved the value of the
  shock viscosity coefficient does not change the amount of heating;
  rather, it only changes the volume (number of grid cells) over which the
  shock energy is spread. For the model shown in this work, we used $c_{\rm sh}=10$.
  
\begin{figure}
  \begin{center}
    \resizebox{\columnwidth}{!}{\includegraphics{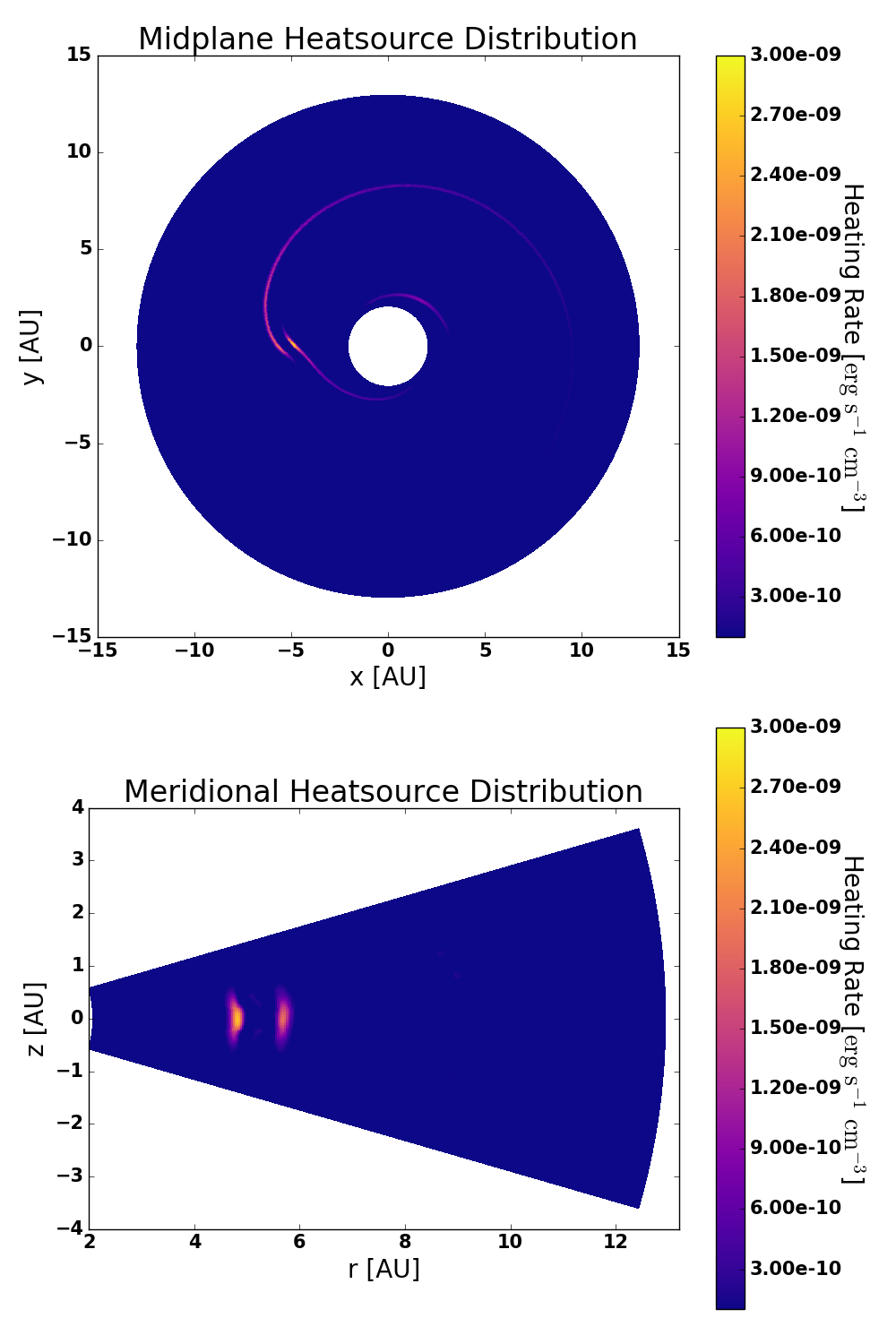}}
\end{center}
\caption[]{The midplane (top) and meridional plane
  containing the planet (bottom), showing the shock heating
  rate distributions. Notice the spiral concentration of the shock heating rate in the midplane and the two lobes of
shock heating on either side of the planet at 5.2 AU.}
\label{fig:shock_heating_rate}
\end{figure}

The shock heating rate is converted from code units into
\mm{cgs units}
for use in RADMC-3D. 
The midplane and meridional resulting shock heating rates are shown in 
\fig{fig:shock_heating_rate}. As a result of testing done for the present work, the
public version of RADMC-3D has been updated to include 
\mm{heated gas}
in every cell as a \mm{potential} source of photons. 

The star, made to match the T-Tauri star from \citetalias{Lyra+16},  is placed at the origin. Its radius is 2 $R_\odot$, where
$R_\odot$ is the solar radius; the star has effective temperature set
to  4,000\,K. HD 100546 is a Herbig Be star, so when deriving synthetic
images we scale the distance to match the higher luminosity, as 
\mm{described} 
in \sect{sect:results}. 
A typical Monte Carlo radiative transfer simulation done for this work has $10^7$ photons. While more photons would yield 
greater resolution, the inclusion of an internal heating source, deep
in the optically thick midplane, slows the radiative transfer computations considerably. 

We note that the standard algorithm of \cite{BjorkmanWood01} 
\ed{assumes} thermal equilibrium. This is not a valid
approximation in the optically thick interior, where a Lagrangian fluid element is suddenly shock heated,
then cools, \ed{but does not reach thermal equilibrium} before being
shocked again. The postprocessing approach nevertheless gives a better
estimate of the temperature in the optically thin regions of the disk
than done with the approximation used in the hydrodynamical model of
\citetalias{Lyra+16}. 


\section{Results and Discussion}
\label{sect:results}

\begin{figure*}
  \center{
    \resizebox{\columnwidth}{!}{\includegraphics{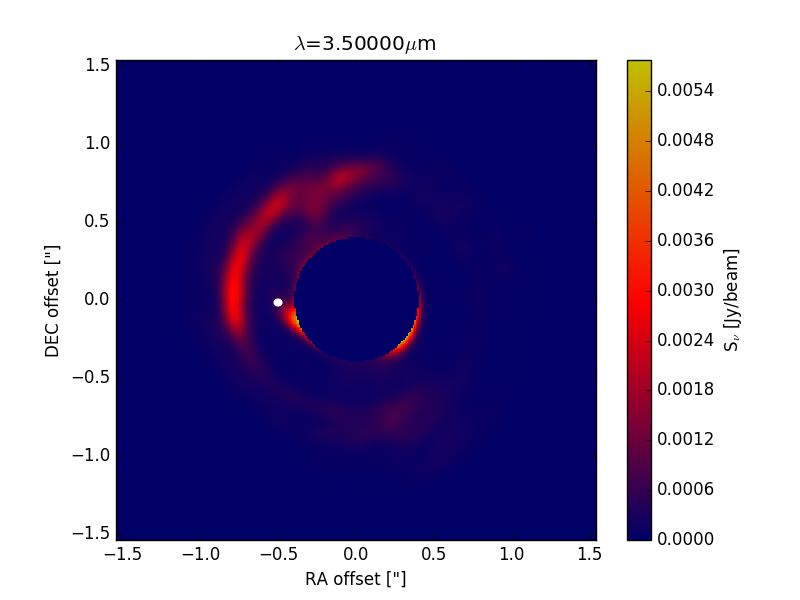}}
    \resizebox{\columnwidth}{!}{\includegraphics{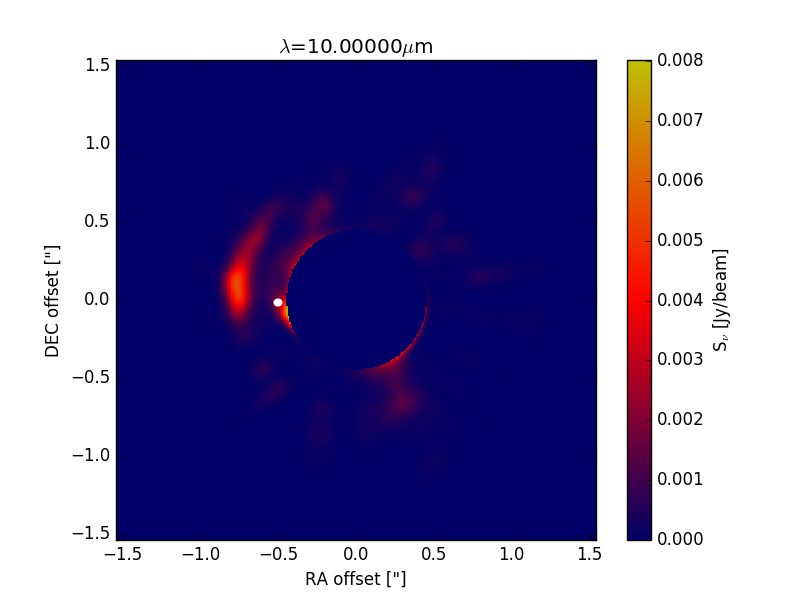}}}
\caption[]{Synthetic disk images with original heating rate and
  including scattering at
  wavelengths of 3.5$\mu$m (left) and 10$\mu$m (right). The white dots
  are the locations of the planet in the midplane. \ed{The coronographic
  mask size is 0.4'' for the 3.5$\mu$m image and 0.45'' for the
  10$\mu$m image.}}
\label{fig:3.5scat}
\end{figure*}

In order to compare our synthetic images to those of HD
100546 from \cite{Currie+14}, both had to be on the same
scale. \citetalias{Lyra+16} assumed a T-Tauri star, so the disk
temperature at 5~AU is 
\mm{of} the same order of magnitude \ed{as} that of a Herbig 
Be star's disk a little inward from 50 AU; the factor 100 due to
distance is mitigated by the factor $>$30 increase in luminosity. 
To match the relative sizes in the image, they are simulated as taken from a distance of 
10 pc, compared to HD 100546 being 100 pc away \citep{Berriman+94}. This maps the position of 5 AU to 
the position of 50 AU in HD 100546 in arcseconds. 

The images are given the same FWHM of 
0.1". A coronographic mask is applied, of radius 0.40". This gives the mask a diameter of 0.80", while the diameter of
the mask in \cite{Currie+14} is approximately 0.70". This small discrepancy can be accounted for by the 
uncertainty in the estimate of scaling the map from 5 AU to 50 AU. The
images at 10~$\mu$m 
\mm{use a} coronographic mask of radius
0.45''. 

\begin{figure*}
  \center{
    \resizebox{.75\columnwidth}{!}{\includegraphics{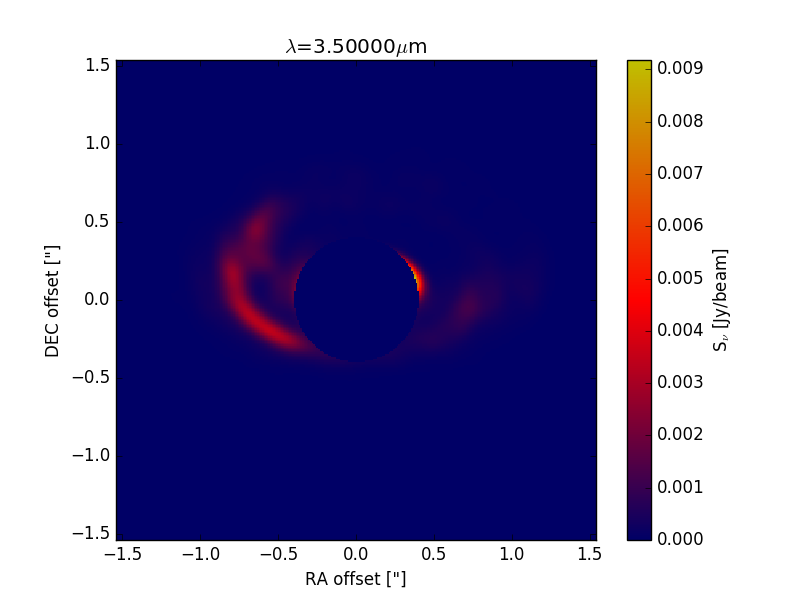}}
    \resizebox{.75\columnwidth}{!}{\includegraphics{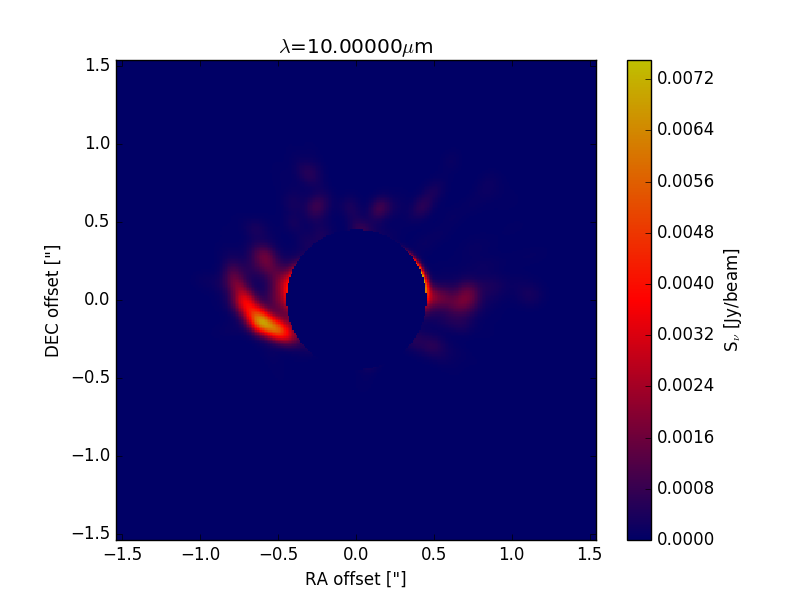}}
    \resizebox{.45\columnwidth}{!}{\includegraphics{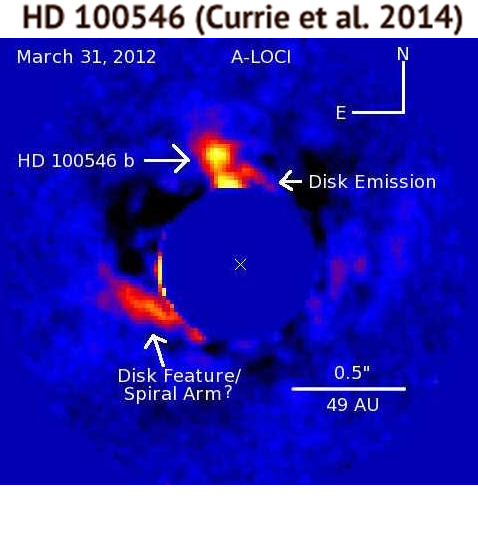}}}
\caption[]{Synthetic disk images with original heating rate and
  including scattered starlight at
  wavelengths of 3.5~$\mu$m (left) and 10~$\mu$m (center), with a 50\degreep~
  inclination angle and a 138\degreep~position angle. On the right is
  the \ed{$L^\prime$ band} \mm{(3.4--4.1~$\mu$m)} observation from \cite{Currie+14}, with the same inclination and
position angles. The location of the planet in the model is just behind the
coronographic mask \ed{between 7 and 8 o'clock}. }
\label{fig:3.5scat_pos}
\end{figure*}

The \mm{face-on} synthetic images at 3.5 and 10~$\mu$m are shown in
\fig{fig:3.5scat}.
\fig{fig:3.5scat_pos} shows the
same images at 50$^\circ$ inclination and 138$^\circ$ position angle,
the viewing perspective of HD\,100546. 

\cite{Currie+14} observed 
unexplained emission around HD 100546, which was not at the 
location of the planet HD 100546 b. It has been proposed that the 
emission is evidence for another planet, HD 100546 c \citep{Currie+15}. The images in \fig{fig:3.5scat_pos} match in general
morphology the observations of \cite{Currie+14}. As evidenced in
\fig{fig:3.5scat}, the spiral
feature in this case is not spiral in the sense of a feature that traverses
different orbital \ed{radii}, but in fact an arc at the same orbital
radius, viewed at an oblique angle. 

\subsection{The effect of shocks alone}

\begin{figure*}
  \begin{center}
    \resizebox{.4\textwidth}{!}{\includegraphics{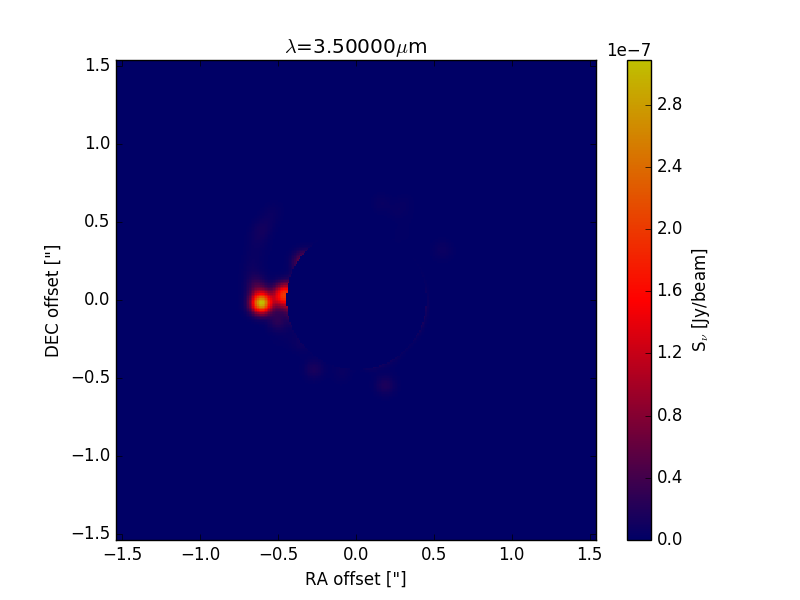}}
    \resizebox{.4\textwidth}{!}{\includegraphics{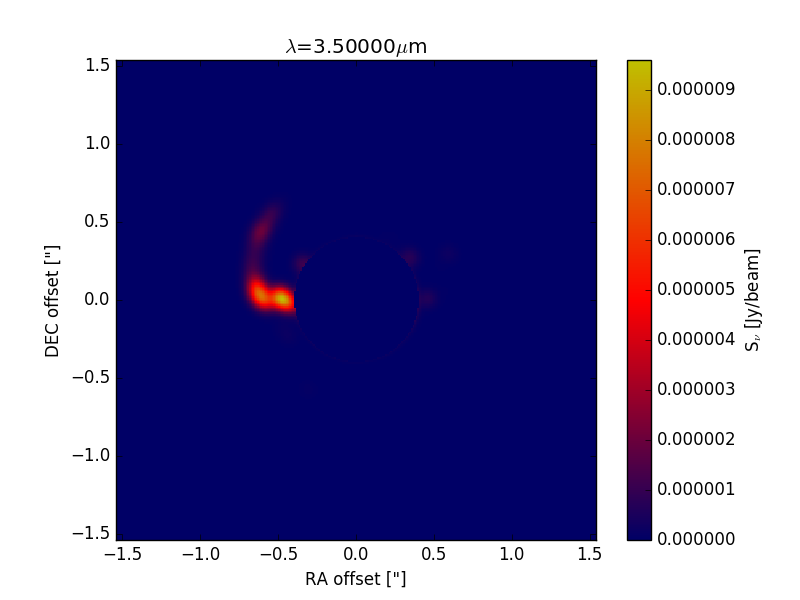}}
    \resizebox{.33\textwidth}{!}{\includegraphics{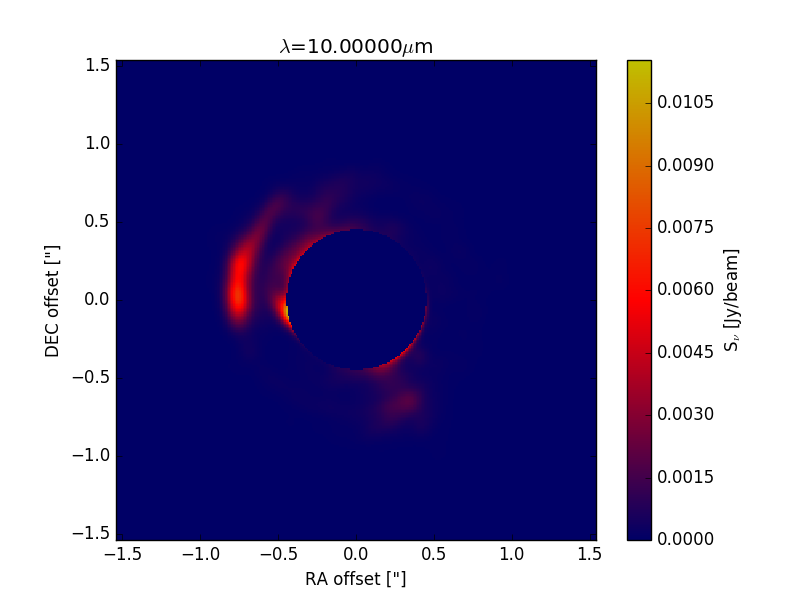}}
    \resizebox{.33\textwidth}{!}{\includegraphics{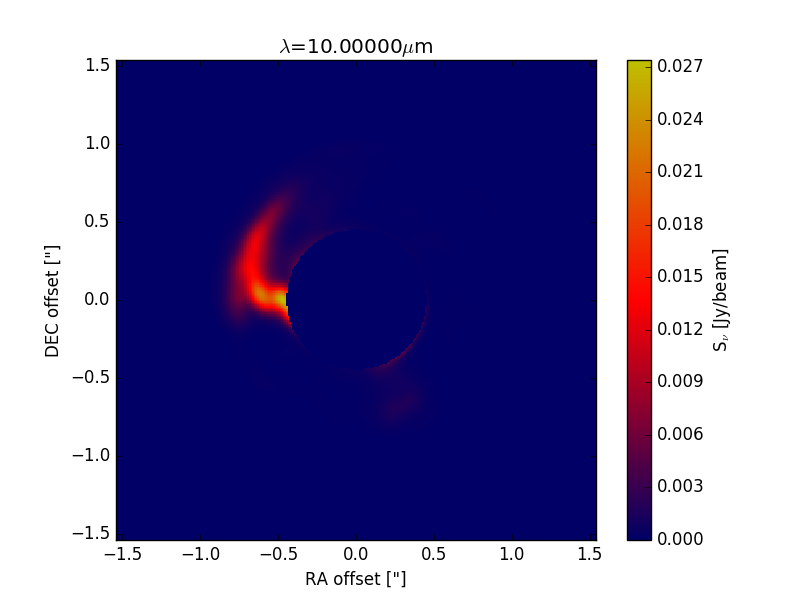}}
    \resizebox{.33\textwidth}{!}{\includegraphics{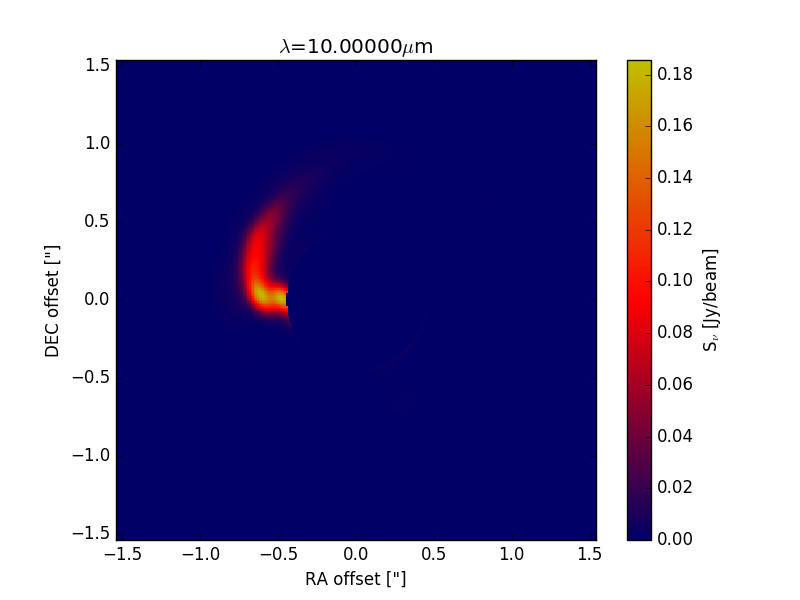}}
\end{center}
\caption[]{Synthetic images generated by the ray-tracing function of
  RADMC-3D, neglecting scattering.
\ed{Two high intensity regions} are seen at a 
factor 10 increase in the shock heating rate (upper left) at a wavelength
of 3.5~$\mu$m, while \ed{the primary spiral wave
is seen} at a factor 20 increase \ed{to the heating rate} (upper right). 
\ed{At a
wavelength of 10~$\mu$m, an arc is seen at an original heating rate
(lower left), as well as more prominently when the heating rate is
increased by a factor of 10 (lower center) and a factor of 20 (lower
right)}. The images are made with 0\degreep~inclination
and position angles. Notice the difference in scale between these
images and those of \fig{fig:3.5scat}. The emission from shocks alone is
orders of magnitude dimmer than the total emission \ed{in the
  3.5~$\mu$m images}. We conclude that
scattering is the main source of emission \mm{at 3.5~$\mu$m}.}
\label{fig:results-images}
\end{figure*}

In order to isolate the emission coming from the planetary shocks, we
calculate a set of synthetic images artificially removing the
scattering opacity from the ray-tracing Monte Carlo simulation. These
are shown in \fig{fig:results-images}. The upper and lower images are calculated
at 3.5\,$\mu$m and 10\,$\mu$m, respectively.
 The images on
the left and right-hand \ed{side of the upper images} have the shock heating rate artificially increased
by a factor 10 and 20, respectively\ed{, while the lower images have
the original shock heating rate, an increase by a factor of 10, and an
increase by a factor of 20, respectively}. 

There is a distinct spiral feature in the infrared that appears more prominently when the shock heating rate is
increased by larger factors. While the image with the original shock
heating rate shows no signs
of the spiral shock in the $L^\prime$ band, the Lindblad lobes are
clearly defined with an increase 
\mm{in the shock heating rate by a} factor \mm{of} 10. At a factor \mm{of} 20, both
the Lindblad lobes and the spiral density wave generated by the planet
are 
\mm{evident}.

When modeled with the inclination and position angles of HD 100546, 
the synthetic observation (\fig{fig:synthetic_image}, left) 
with a factor 20 increase to the heating rate \mm{does} include a spiral feature that matches
that of the observation. Images at longer wavelengths, from
8-15$\mu$m, show even clearer
spirals with \ed{no increase to the shock heating rate}. A synthetic image made at
10$\mu$m with the {\it original} shock heating rate \mm{but no
  scattering} is shown \ed{at the same orientation as HD 100546} in 
\fig{fig:synthetic_image}, right \ed{panel}. \ed{It is almost
  identical to the 10$\mu$m image that includes scattered light
  (\fig{fig:3.5scat_pos}, center), meaning that the emission at
  10$\mu$m is mostly thermal.} The spiral feature shown in the
left of the image \mm{again does} match the morphology of the feature in the
\cite{Currie+14} image.

The similarity in the thermal images suggests \mm{in principle} that the source of the disk
feature 
\mm{could} be a high mass planet. 
However, 
\mm{this image is both much fainter than the scattered light image
  (note the intensity scale in each), and}
required an \textit{ad hoc} factor \mm{of} 20 \mm{increase of the
  shock heating rate} to reproduce
the \mm{{\it $L^\prime$}} observation.

\section{\ed{Pinpointing the source of the emission}}
\label{sect:discussion}
 \mm{We explore in this section the origin of the emission in
our models.}
\cite{Currie+14} finds that the emission feature in HD 100546 shows
little polarization. This was taken as an indication that the emission
must be thermal, and not scattered radiation.  \ed{
\mm{However, an}  error in the GPI pipeline 
made the feature seen in
\cite{Currie+14} appear weakly polarized.  In fact the emission should
be interpreted as strongly polarized \mm{Millar-Blanchaer (2017, priv.\
  comm.)}} 

\begin{figure*}
  \begin{center}
    \resizebox{0.4\textwidth}{!}{\includegraphics{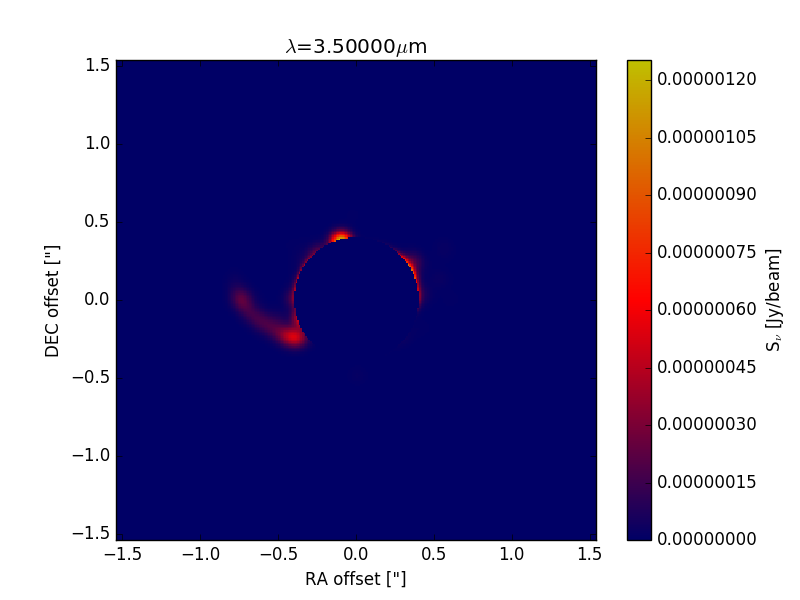}}
    \resizebox{0.4\textwidth}{!}{\includegraphics{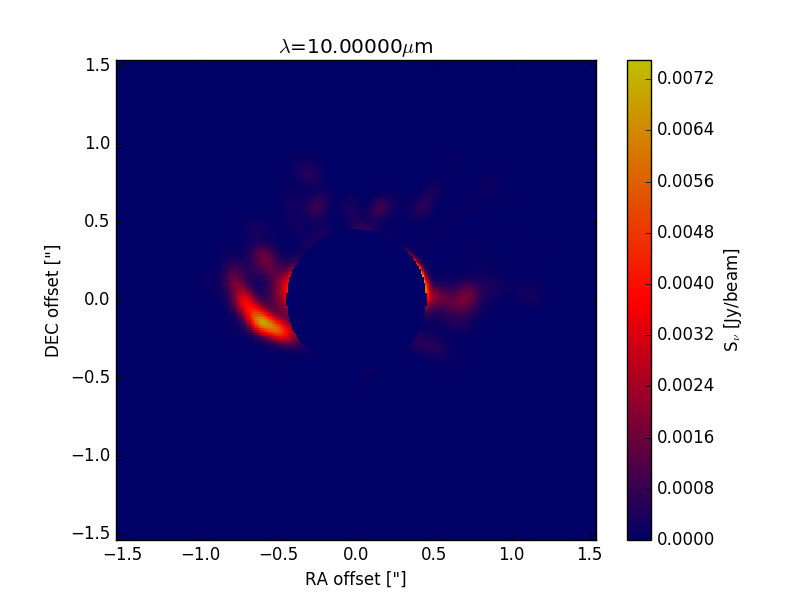}}
  \end{center}
\caption[]{Synthetic images without scattering with a 50\degreep\
  inclination angle and 138\degreep\ position angle. The 3.5~$\mu$m
  image (left) was created with a factor 20 increase to its shock
  heating rate. The 10~$\mu$m image (right) was created using the
  original shock heating rate.}
\label{fig:synthetic_image}
\end{figure*}

\mm{As} we showed,
a match in morphology with our 
\mm{3.5~$\mu$m} images requires an arbitrary
increase in the magnitude of the shock heating rate of a factor 20,
which may not be reasonable. Conversely, the spatial position of the
intensity feature in our model \ed{(\fig{fig:3.5scat}) matches not the
immediate vicinity of the planet, as expected if it came from the
hot Lindblad lobes, but from the outer gap wall. Between 7 and 9 AU there is an
increase in the disk scale height, puffing it upwards. It is this feature that produces
the emission seen in \fig{fig:3.5scat} and \fig{fig:3.5scat_pos}}. This outer gap
reaches above the average disk height, and scatters photons from
the central star towards the observer. \ed{\fig{fig:3.5scat} also shows
that the contribution from shocks at 3.5~$\mu$m is negligible
\mm{compared to the scattered light}. 
\mm{Thus} the emission in our 
model is not primarily thermal, but mostly polarized scattered light.
On the other hand, at wavelengths of 10~$\mu$m or longer, thermal
emission comes to dominate. In fact, a spiral feature matching the
observation of HD 100546 is present in both the 10~$\mu$m image including
scattered light (\fig{fig:3.5scat_pos}, center) and including only
thermal emission (\fig{fig:synthetic_image}, right).} 

\begin{figure*}
  \begin{center}
    \resizebox{.49\textwidth}{!}{\includegraphics{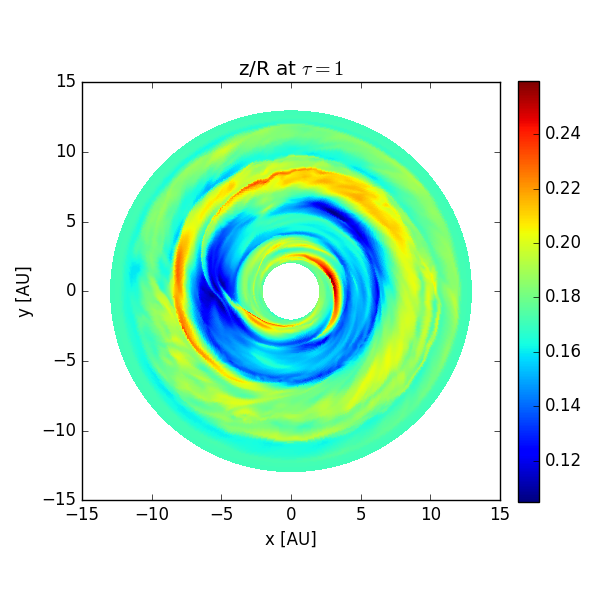}}
    \resizebox{.49\textwidth}{!}{\includegraphics{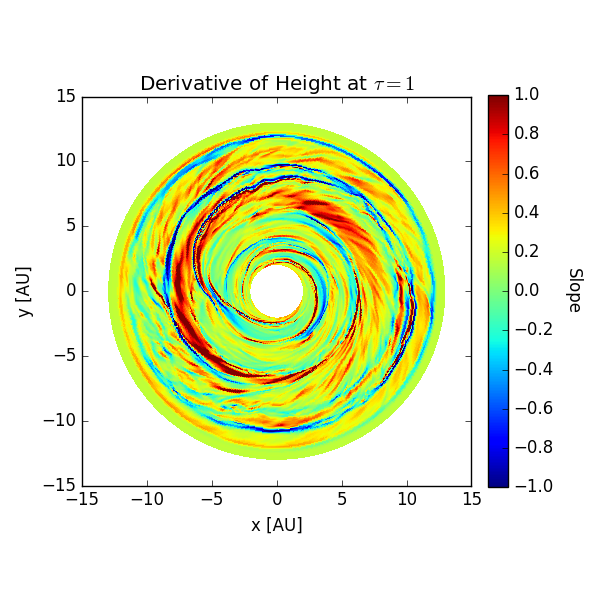}}
  \end{center}
  \caption[]{{\ed {\it Left.} Normalized height $z/R$ at the surface of
    optical depth $\tau=1$. In this plot, the highest point at a given
    angle should be prominent in scattered light, while what is behind
    it should be shadowed.  {\it Right.} The derivative of the height,
    $dz/dR$. Positive slopes 
    \mm{tilt} toward the star, while negative slopes
     \mm{tilt} away, and thus are in shadow. Not all positive slopes should appear in
    scattering, because some may be shadowed by features further in. }}
\label{fig:tauone_temp_dens}
\end{figure*}

\mm{What determines the morphology of the scattered light image?}
\ed{We plot in \fig{fig:tauone_temp_dens} 
the normalized
height of the $\tau=1$ surface $z/R$ (left panel), and the derivative of the height, $dz/dR$ (right panel). In the
left panel the features with positive slope tilt toward the
star and should appear in scattered light \mm{if they are not shadowed
  by features closer to the star}, while features with negative
slope should be shadowed. 
The highest 
\mm{feature at any} given angle will
appear in scattered light. Compare this image to the left plot of
\fig{fig:3.5scat}. 
\ed{With the right side of the image at 0$\degree$, increasing
  counter-clockwise,} the region in the third quadrant between \ed{270$\degree$
and 165$\degree$} in the outer disk is in the shadow of the inner disk. So is 
the region of the first and fourth quadrant between \ed{75$\degree$ and 330$\degree$.} 
In \fig{fig:3.5scat}  we can see some of the inner disk scattering, 
but it is too close to the coronographic
mask. }

\begin{figure*}
  \begin{center}
    \resizebox{.545\textwidth}{!}{\includegraphics{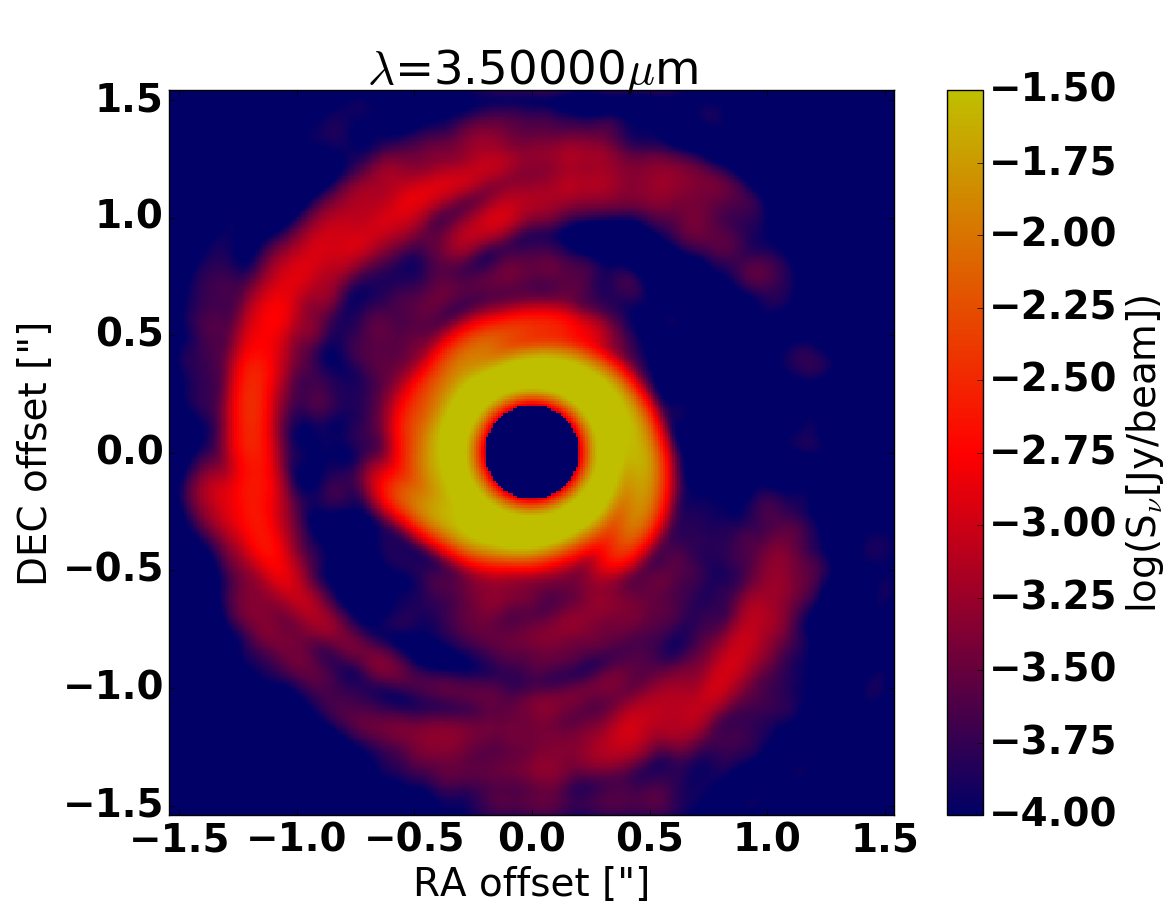}}
   \resizebox{.445\textwidth}{!}{\includegraphics{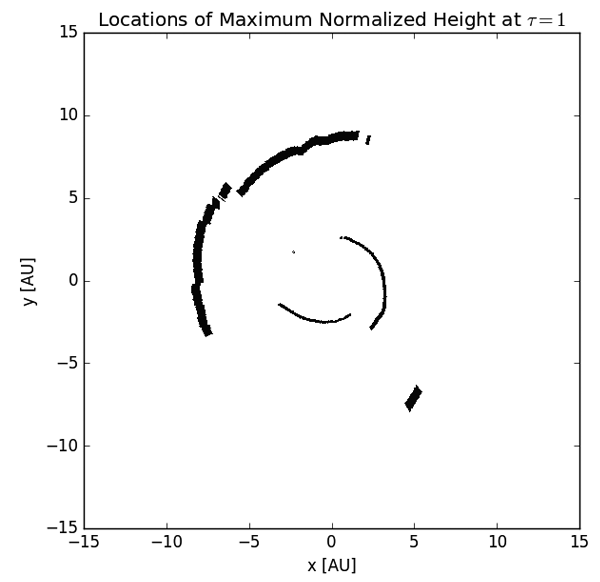}}
  \end{center}
  \caption[]{\ed{{\it Left.} Same as the left panel of \fig{fig:3.5scat}, but with a
    smaller coronographic mask (0.2''), showing the inner spirals, and
    plotted in logarithm. The inner spirals seen in
    \fig{fig:tauone_temp_dens} shadow the outer disk. {\it Right.} The
  radial location of the maximum normalized height at a given angle. The
  surface it traces matches the prominent emission features in \fig{fig:3.5scat}.}}
\label{fig:reducedmask}
\end{figure*}

Figure~\ref{fig:reducedmask}, left panel, shows the same 
\mm{simulation} as the left panel of 
\fig{fig:3.5scat}, but with the coronographic mask reduced to 0.2'',
and 
\mm{a logarithmic color bar.} The inner spirals are bright, so
the high areas behind them in the outer disk are in shadow. In the right
panel of \fig{fig:reducedmask} we plot the radial
location of the maximum normalized height at every angle. The surface it traces is a
good match to \fig{fig:3.5scat}.

Therefore, the match in morphology shown in the
previous section between our synthetic image including scattered light
and the image from \cite{Currie+14} \mm{can indeed be interpreted to show} how a high-mass planet
\mm{could} 
produce an observational signature similar to that in the HD
100546 system.


\begin{figure*}
  \begin{center}
    \resizebox{.49\textwidth}{!}{\includegraphics{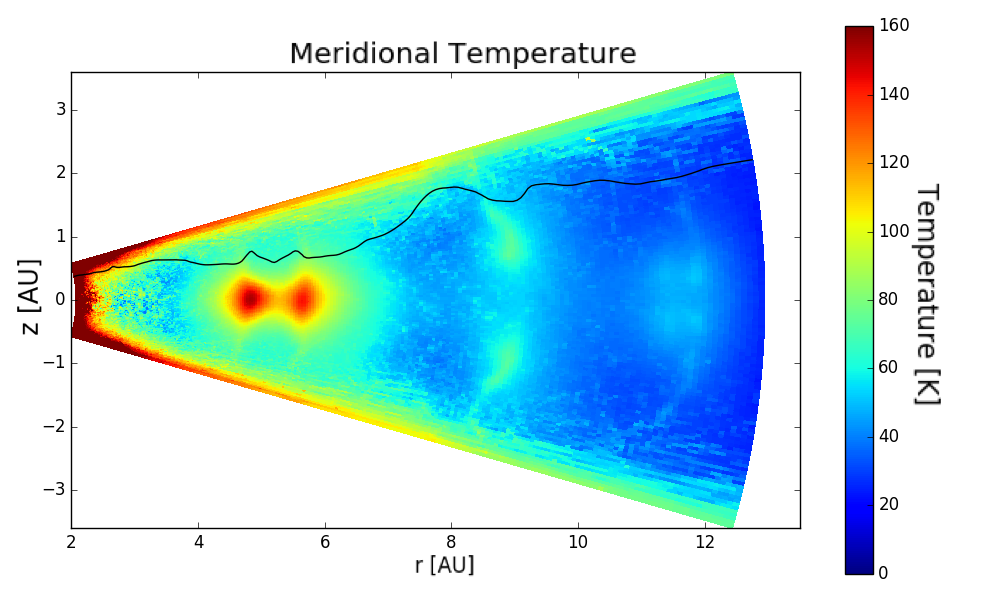}}
    \resizebox{.49\textwidth}{!}{\includegraphics{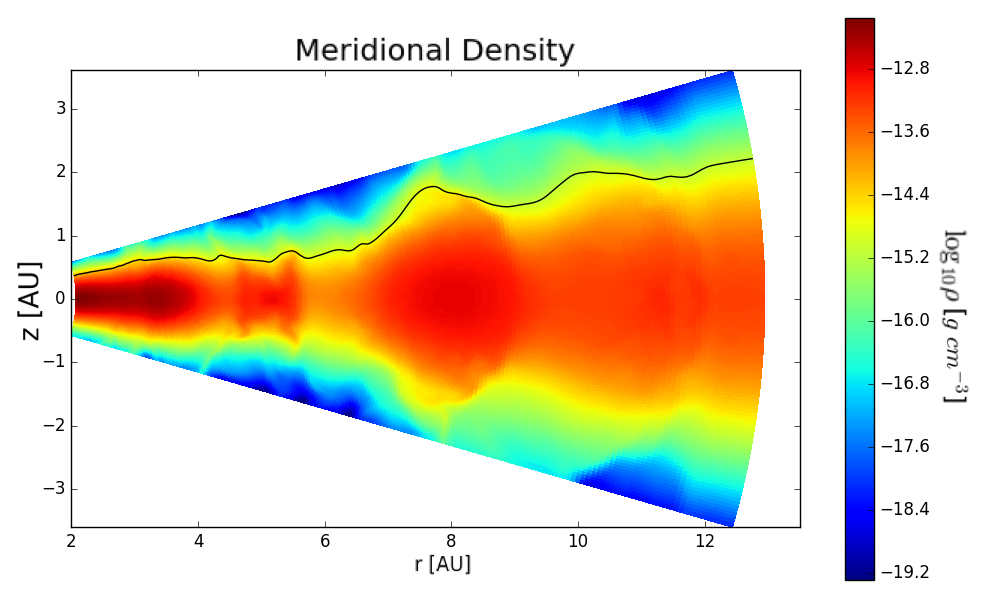}}
  \end{center}
  \caption[]{Meridional temperature (left) and density (right) in the plane of the planet, with a
  contour drawn in black for the surface of $\tau=1$ at a wavelength of
  3.5$\mu$m. This has no increase to the shock heating rate and is the
  result of a RADMC3D thermal Monte Carlo run with $10^8$ photons.
  \ed{The emission is not primarily from the thermal shocks but mostly
  from scattering off the outer gap wall between 7 and 9
  AU that protrudes above the average disk scale height.}} 
\label{fig:meridional_temp_dens}
\end{figure*} 

\begin{figure*}
 \begin{center}
   \resizebox{\textwidth}{!}{\includegraphics{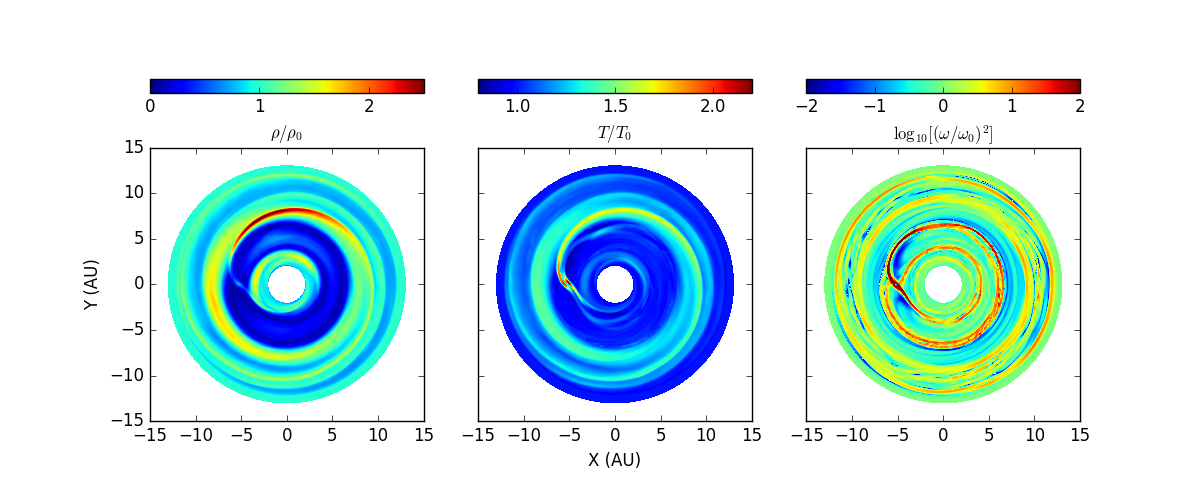}}
 \end{center}
 \caption[]{State of the disk midplane: density (left), temperature
   (middle), and vorticity (right). \ed{The density shows a feature in the
   outer edge of the gap, with peak density between 8 and 9 o'clock.
   The temperature plot shows that the outer gap edge is hotter than the gas around it. The vorticity
   does not particularly show a strong vortex, although the density
   maximum stands on a vorticity depression}.}
\label{fig:midplane-quantities}
\end{figure*}


\subsection{\ed{Ruling out a vortex}}

\begin{figure*}
  \begin{center}
    \resizebox{\textwidth}{!}{\includegraphics{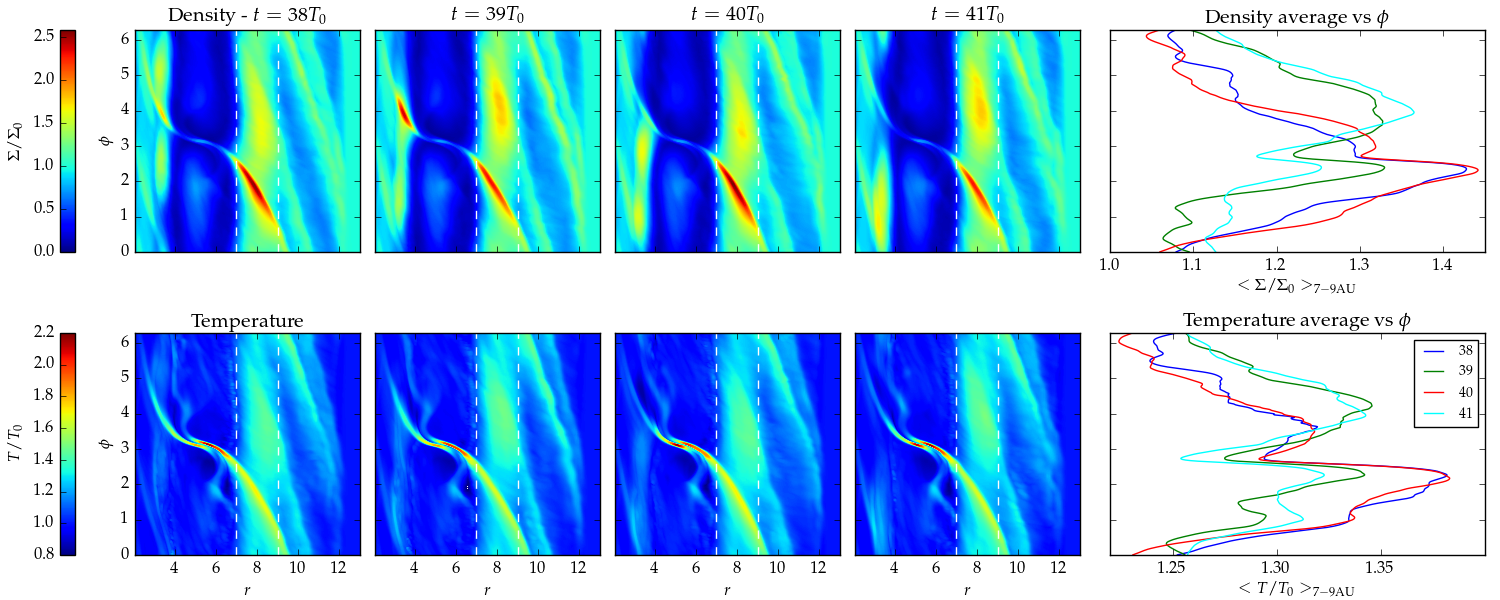}}
  \end{center}
  \caption[]{Density (upper plots) and temperature (lower plots) at
    the midplane in four snapshot\ed{s} separated by one planetary orbit. The
    rightmost plot shows the average in the region between 7 and 9 AU
    as a function of azimuth. We see that the \ed{high density} feature is
    stationary with respect to the planet. A vortex is an independent
    entity and would show synodic motion, so we rule out
    the vortex \ed{possibility}. However, the radial averages show
    periodic variation at the synodic period (2 planetary orbits), at
    the 10\% level. We conclude that even though a vortex may be
    present, \ed{ the corotating feature is dominant}. The maxima of density and temperature 
    in the second quadrant is due to the presence of the primary
    spiral arms launched at Lindblad resonances.}
\label{fig:dens-temp-backforth}
\end{figure*}

\begin{figure*}
  \begin{center}
    \resizebox{.245\textwidth}{!}{\includegraphics{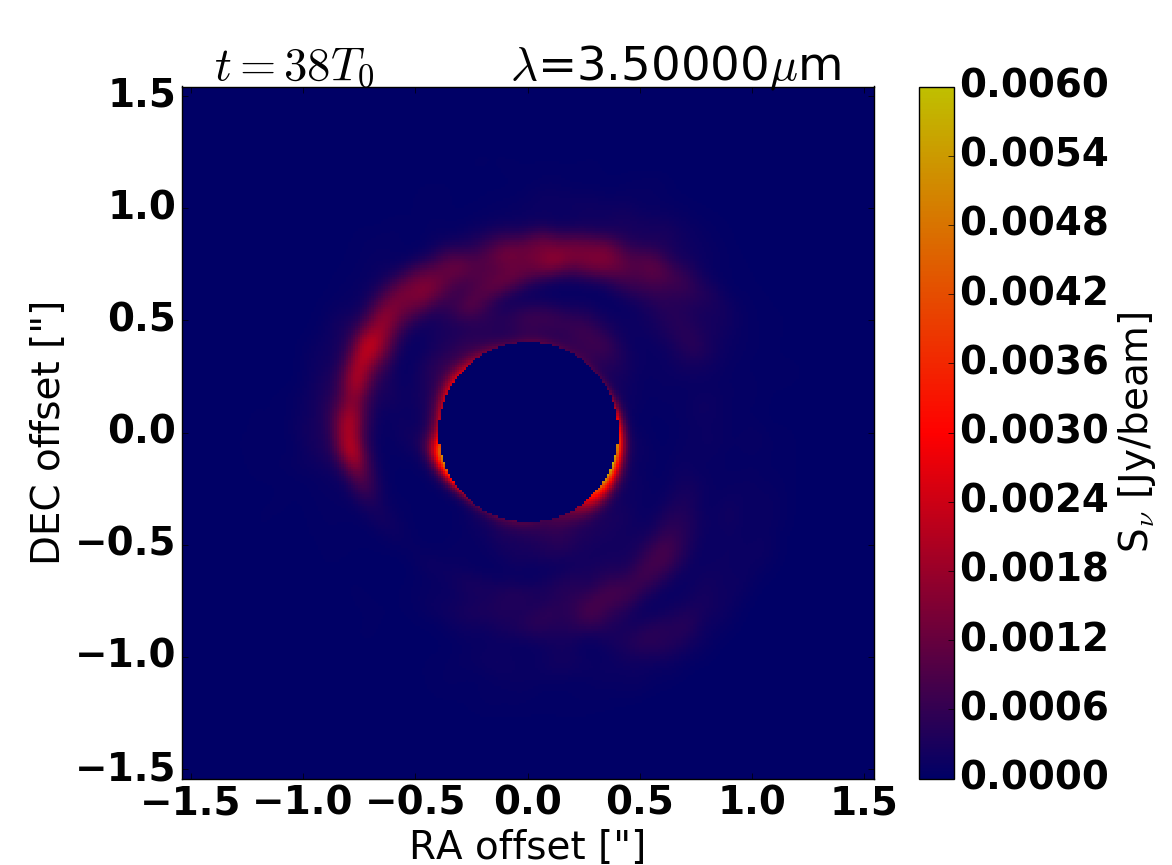}}
    \resizebox{.245\textwidth}{!}{\includegraphics{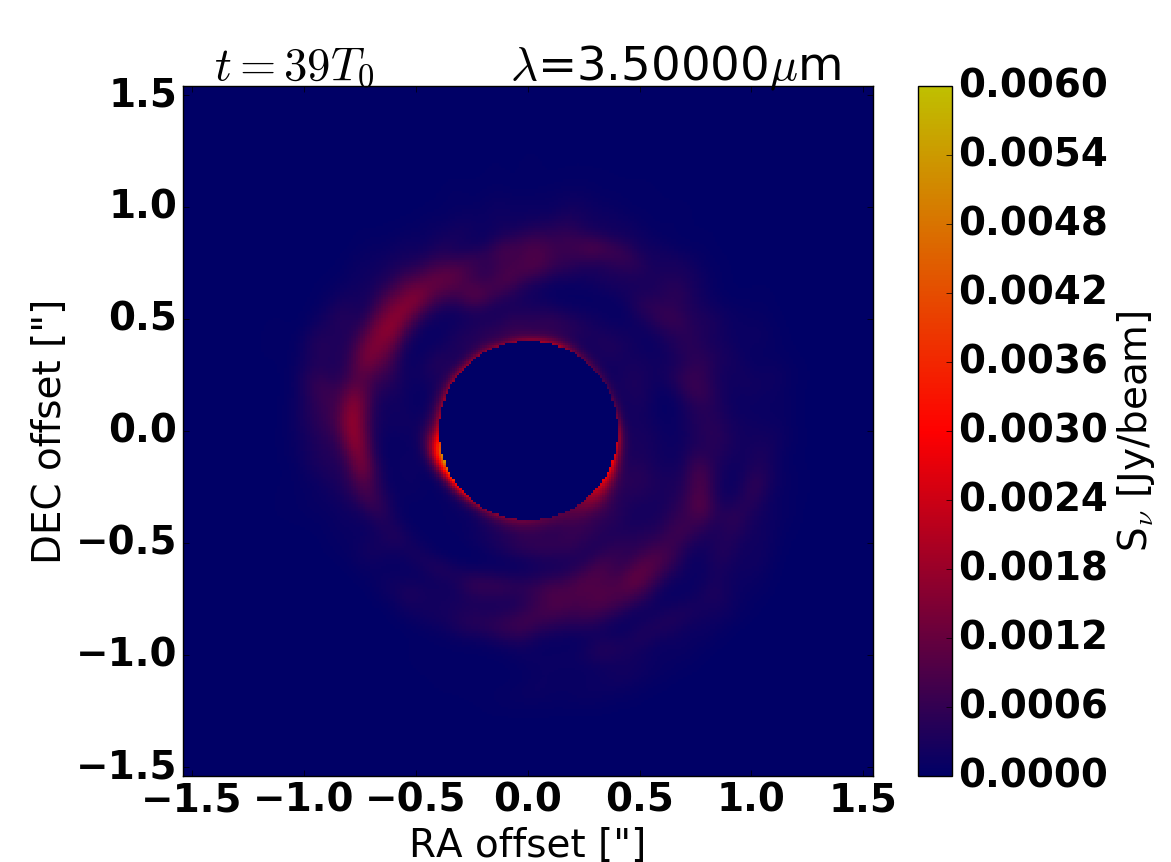}}
    \resizebox{.245\textwidth}{!}{\includegraphics{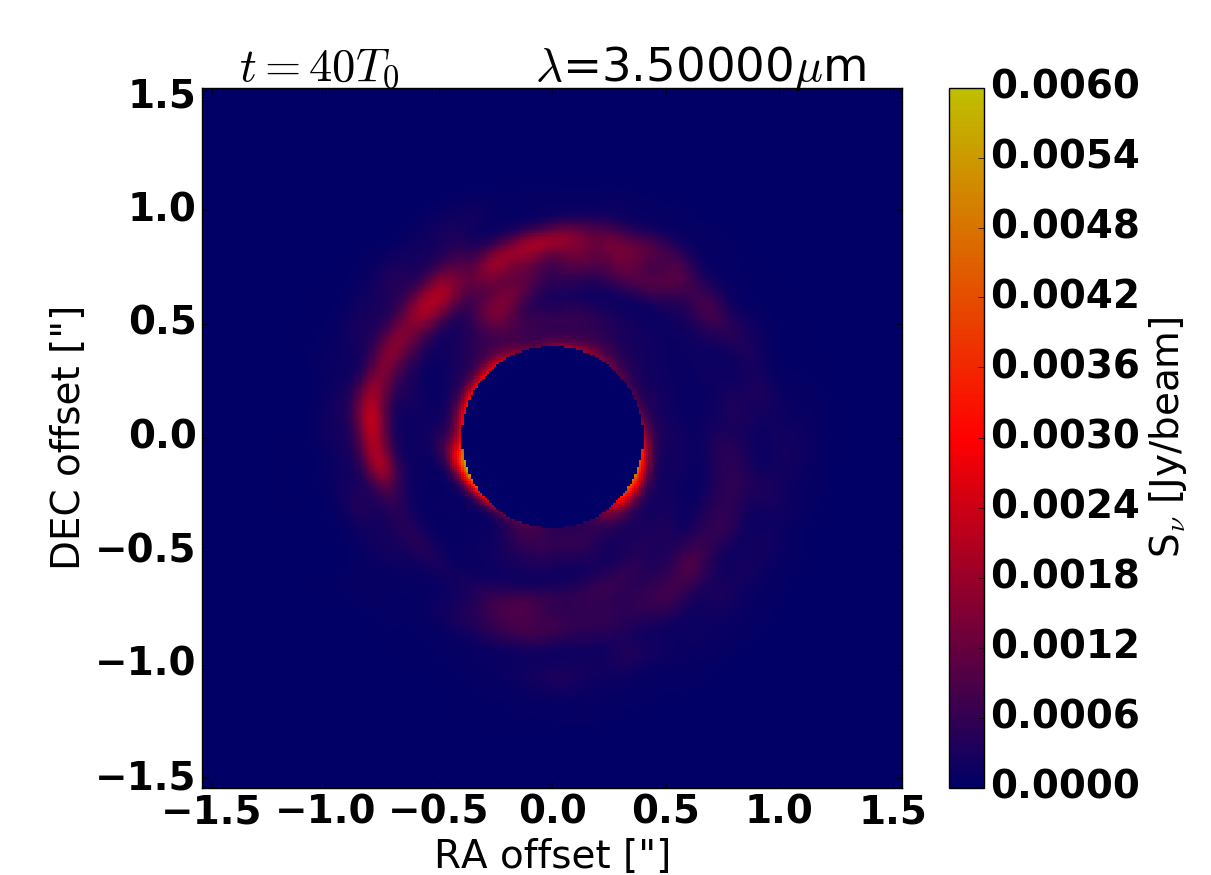}}
    \resizebox{.245\textwidth}{!}{\includegraphics{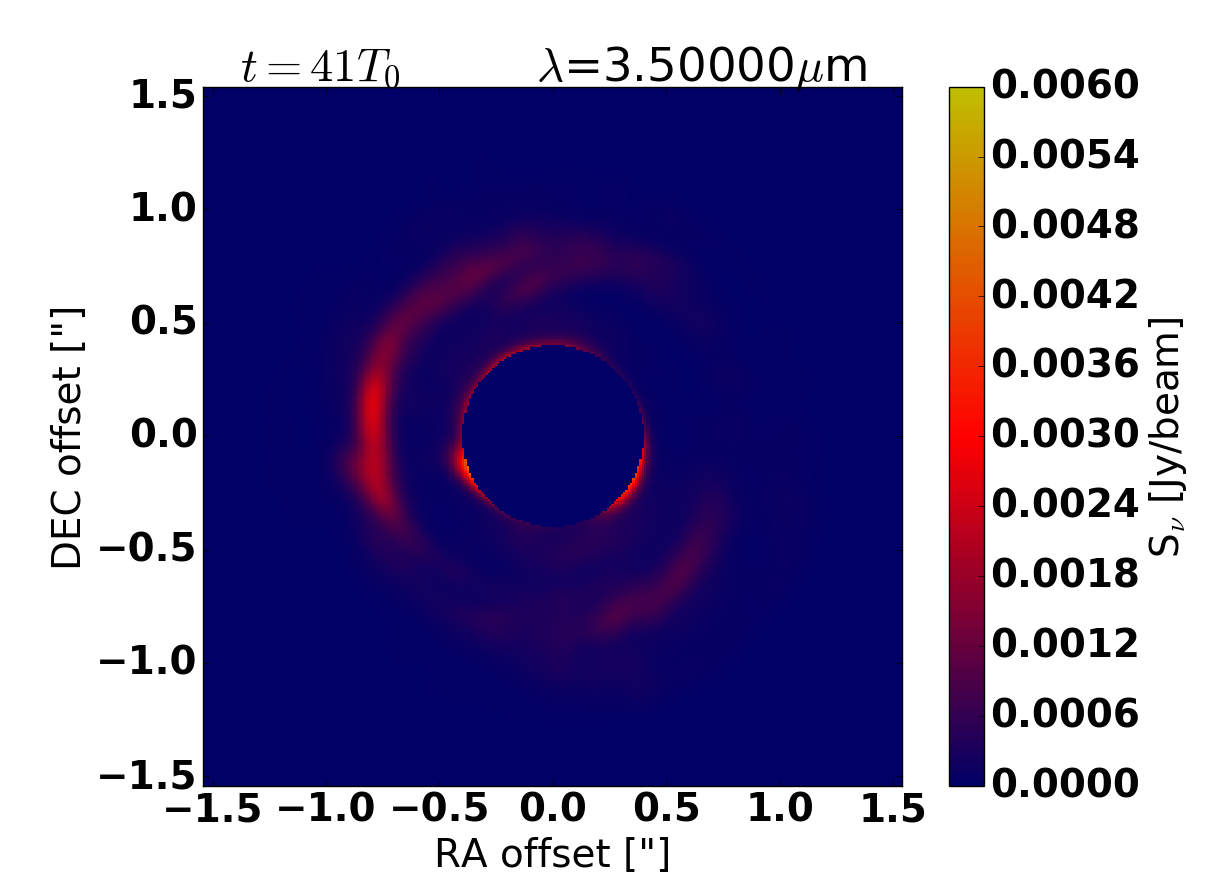}}
  \end{center}
  \caption[]{Scattering image in the same snapshots as
    \fig{fig:dens-temp-backforth}. The intensity maxima occur outside
    the gap and mostly in the second quadrant. The high density
    features in the snapshots are mostly located in the third quadrant, so they are not responsible for the emission.}
\label{fig:scattering-images}
\end{figure*} 

The bright feature \mm{in the second quadrant} is seen in
scattered light, not thermal emission, and results from an enhanced
scale height immediately outward of the
planet. The \ed{enhancement} is asymmetric, lifting the side next to the
planet more than the side opposite to it. 

\ed{\mm{One type of} hydrodynamical feature that produces asymmetry
  in disks is a vortex \citep{LyraLin13,vanderMarel+13}, which is
  expected to form at planetary gap edges as a result of the
Rossby wave instability \citep{Lovelace+99,deValBorro+06,Lyra+09b}.} 
We plot in \fig{fig:midplane-quantities} the midplane values of
density, temperature, and vorticity. The temperature is that calculated in the original Pencil Code
simulation, not the MC
computation. There is a noticeable density increase and
temperature increase associated with the location of the bright
feature, next to the planet. \ed{There is 
an enhancement in density at the
outer gap wall, near opposition, between 8 and 9 o'clock, which could be a vortex. The whole outer gap wall
is a vorticity depression (\fig{fig:midplane-quantities}, right
panel), as expected, but a clear localized vorticity minimum is not associated
with the density maximum.}


To see if there is a correlation between the \ed{high density}
feature and the scattering
images, we plot in \fig{fig:dens-temp-backforth} four snapshots of the
disk, from 38 to 41 orbits of the planet, spaced one orbit
apart. The upper panels show the density, the lower panels the
temperature. The rightmost plots show the density and
temperature as a function of azimuth, averaged in the region from 7 to
9 AU, bracketed by the dashed lines in the contour plots. \mm{The high
  density feature is at $\phi \approx 4$.}

For comparison, we plot the intensity calculated from the respective
snapshots, shown in \fig{fig:scattering-images}. The planet is at
$\phi=\pi$ (at nine o'clock), and azimuth increases counterclockwise. In
the panels of \fig{fig:dens-temp-backforth} the feature is located in the
third quadrant. In the intensity plots of \fig{fig:scattering-images}
the maxima are located in the second quadrant mostly.  The locations
do not match, so we must seek another explanation. 

More intriguingly, notice that the \ed{high density} feature seems stationary in
the reference frame of the planet. An \ed{independent} feature at 8
AU is close to the 
2:1 resonance with a planet at 5 AU, and thus should \ed{alternate
conjunction and opposition with the planet} in snapshots separated by one planetary
orbit. The fact that this does not happen for this feature is clear
indication this is not \ed{an independent feature, so we rule out the
vortex possibility}. Yet, notice that the right plots of
\fig{fig:dens-temp-backforth}  clearly show variation at the 10\% level 
with a period of two orbits, the synodic period. We conclude that the
outer gap edge harbors underlying independent structures, but they are of lower
prominence compared to the overarching stationary pattern.

\subsection{Secondary Spiral}

The fact that the feature, both in the midplane (\fig{fig:dens-temp-backforth}) and in the intensity
plots (\fig{fig:scattering-images}), seems to corotate with the planet brings us to consider the
spiral patterns again. While the primary arm is launched at Lindblad
resonances, \cite{Juhasz+15} and \cite{Zhu+15} call attention to the existence of
secondary spiral features, which \cite{FungDong15} and \cite{Lee16} show are launched at the 2:1 resonance
(similarly, tertiary arms are also present, launched from the 3:1
resonance, and so on). These secondary arms for high-mass planets 
were already noticed by \cite{deValBorro+06}, but their 
origin or nature had not been studied. \cite{FungDong15} and
\cite{Dong+16} explored their
observational properties, finding that they are prominent in scattered
light, and also derive a relationship between the planet mass and the
angular distance between the primary and secondary arms. For a 5$M_J$
planet, they should be separated by $\approx$141$^\circ$. 

\begin{figure}
  \begin{center}
    \resizebox{\columnwidth}{!}{\includegraphics{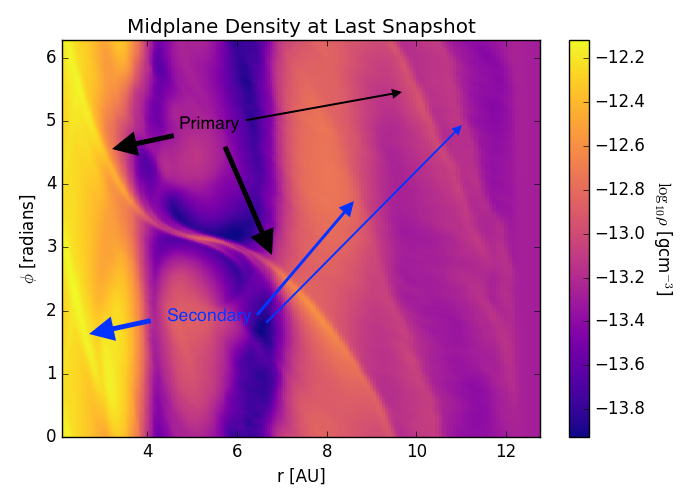}}
  \end{center}
  \caption[]{Density at last snapshot, in polar coordinates,
    showing the primary spiral, launched at the Lindblad resonances,
    and the secondary spiral. The thinner arrows show the periodic
    continuation of each.}
\label{fig:secondary_spiral}
\end{figure}

\begin{figure*}
  \begin{center}
    \resizebox{\textwidth}{!}{\includegraphics{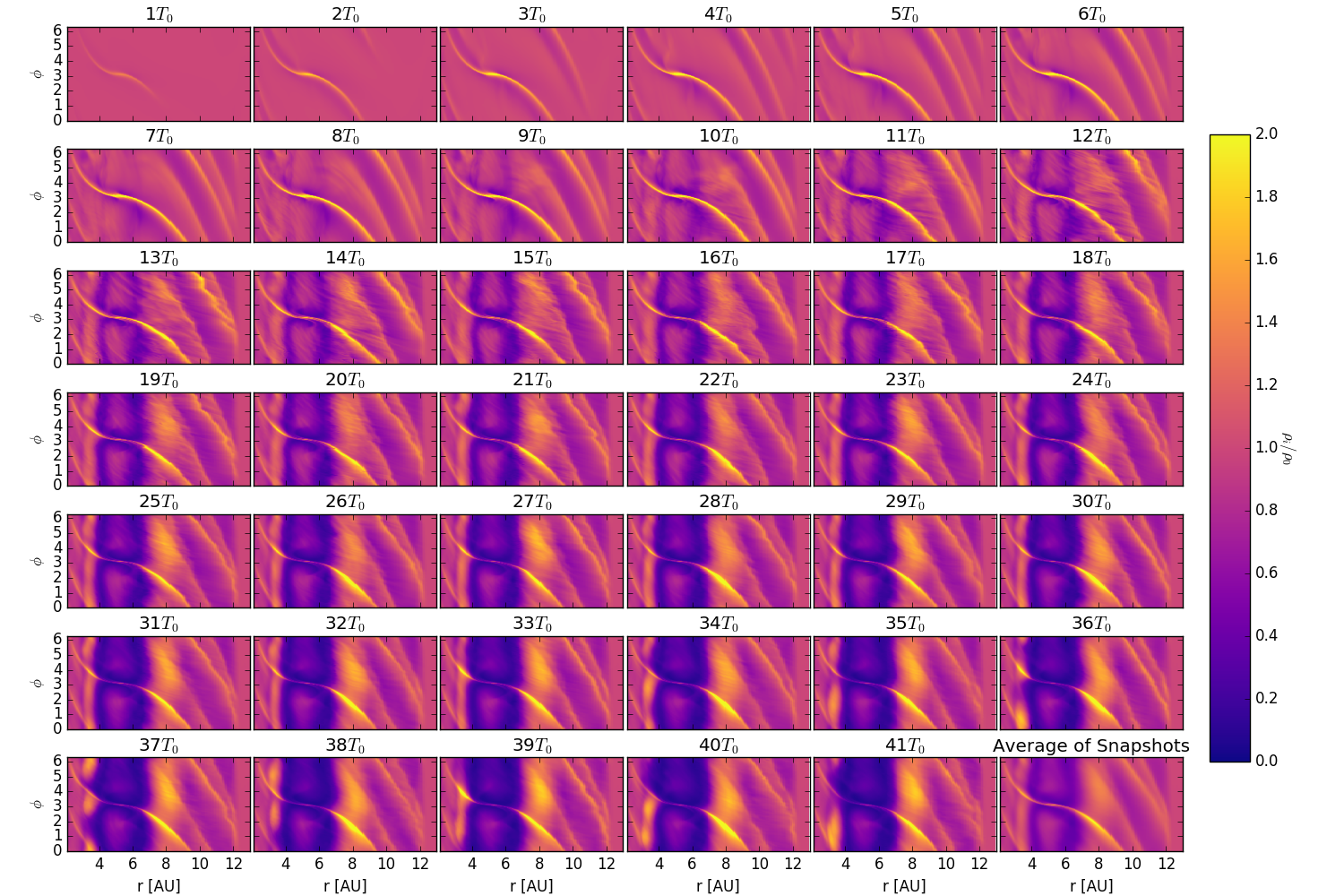}}
  \end{center}
  \caption[]{Density in the midplane in snapshots from one to 41 orbits,
   with one orbit cadence. In addition to the primary Lindblad spiral,
   the planet launches secondary spiral arms from the 2:1 resonance,
   located at $\approx$ 3 and 8 AU. The inner secondary is seen
   already at the second snapshot. The outer secondary spiral is visible from
   the fourth snapshot but more prominently at the sixth. It is seen as an
   enhancement in density \ed{radially inwards from} the rarefaction
   wave of the primary 
   spiral, launched from $r\approx 8$ AU and $\phi \approx 5$.
   The secondary spirals overlap with the forming planetary gap edges as time 
   increases.}
\label{fig:secondary_spiral_timestep}
\end{figure*}

We plot in \fig{fig:secondary_spiral_timestep} snapshots of the disk
midplane, with one planetary orbit cadence. The secondary arm in the
inner disk is visible from the second orbit, and in the outer disk from
the fifth or sixth orbit. \ed{A labeled closeup of the features in the last
snapshot can be seen in \fig{fig:secondary_spiral}.} The outer spiral is seen as a density enhancement
\ed{between the primary spiral and its periodic continuation}, launched from
$r\approx 8$ AU and $\phi$ between $\sim$5--6. 

The \ed{high density} feature seen in \fig{fig:dens-temp-backforth} is obvious in later snapshots and clearly had
its origin in the secondary outer arm. The panel in the lower right is an average
of all snapshots. The measured separation is $\approx$150$\degree$,
within 6\% of the value predicted by the formula of
\cite{FungDong15}. We conclude that indeed we are seeing the
secondary spiral.  




A novel aspect of our model is that we also calculate the temperature, showing
that \ed{the outer gap edge and} the secondary spiral are hotter features. Being so, they
raise the scale height locally and increase the intensity in
the scattered starlight synthetic images. The spiral is a stationary feature
in reference to the planet, which matches the stationary area of high
intensity seen in \fig{fig:scattering-images}. Another difference
between our model and those of \cite{FungDong15} and \cite{Dong+16} is
that we do not 
use an artificial eddy viscosity. The large scales of the flow
are effectively inviscid. Either the lack of viscosity or the
presence of buoyancy, or both, can lead to the discrepancies between
the fuzzy spirals we see versus their well-defined ones. 

\subsubsection{Eccentricity}

\begin{figure}
  \begin{center}
    \resizebox{.5\textwidth}{!}{\includegraphics{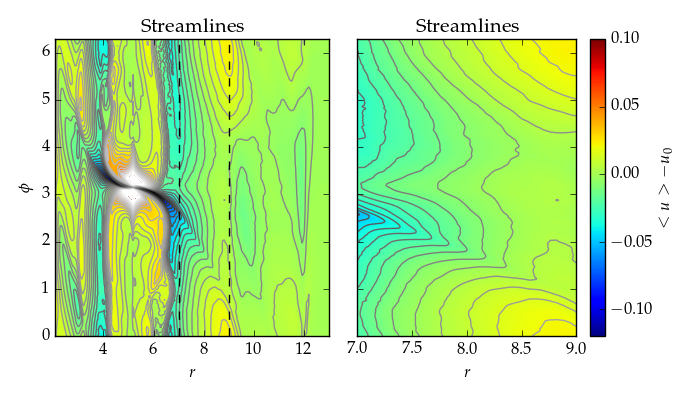}}
  \end{center}
  \caption[]{Flow streamlines. The color code refers to the midplane
    speed, $\sqrt{u_r^2 + u_\phi^2}$, normalized by the initial value
    of this quantity. The right plot zooms in at the region between 7
    and 9 AU. One sees that the flow is slower \ed{at opposition} and
    faster \ed{at conjunction}, as would be expected if the orbits were
    slightly eccentric, with aphelion occurring \ed{at opposition}.}
\label{fig:streamlines}
\end{figure} 

\begin{figure*}
  \begin{center}
    \resizebox{\textwidth}{!}{\includegraphics{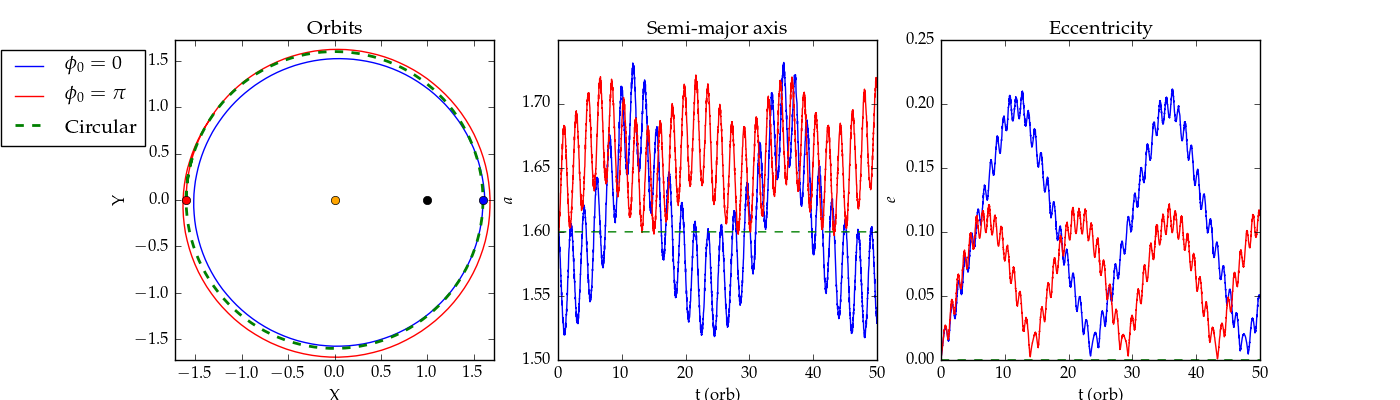}}
  \end{center}
  \caption[]{We place two test particles in the same outer orbit, initially circular, but with the particles started at opposition (blue dot) and at conjunction (red dot). An orange dot and a black dot mark the position of the fixed star and planet, respectively. A circular orbit (green) is shown for comparison. The first two orbits are shown in the leftmost plot (blue and red lines). The particles develop an eccentricity, both having closest approach to the planet (at $\phi=0$) in aphelion. The middle panel shows the semi-major axis and the right panel the eccentricity. The particle at opposition is more heavily perturbed by the planet. Both orbits should change in time as a result of the perturbation.}
\label{fig:eccentricities}
\end{figure*} 

%



Although we identify the secondary spiral as the culprit of the
lopsided emission we see in our models, we have to examine as well to
what degree eccentricity excitation is or is not influencing the
results. \cite{KleyDirksen06} \ed{argue} that a massive planet turns the
disk gap eccentric by removing gas from the 2:1 resonance. This
resonance provides eccentricity damping so, without it, the gap walls
exponentially grow into an eccentric state within some hundreds of years. The
timescale for eccentricity pumping, for a 5$M_J$ planet is 250
orbits, reaching a saturation value of 0.25 in $\approx$700 orbits. 

We assess to what eccentricity our system has grown, if any. If a
system of e-folding time \ed{$\tau=250 T_0$ saturates at $e_{\rm
  sat}=0.25$ at $t_f \approx 700
T_0$, then at $t=40 T_0$ it should have grown to only} 
\begin{eqnarray}
\ed{e} &\ed{=}& \ed{e_{\rm sat}
  \exp\left[-\left(t_f-t\right)/\tau\right]}  \nonumber\\
&=& 0.25\exp\left[-\left(700-40\right)/250\right] \approx 0.02
\end{eqnarray}
We plot in \fig{fig:streamlines} the streamlines of the flow in the
midplane. The color code refers to the reference speed $(u_r^2 +
u_\phi^2)^{1/2}$ normalized by the initial condition. The average is
in time, from 20 to 41 orbits, one snapshot per orbit. The right panel 
zooms in the region from 7 to 9 AU, between the dashed lines. It is
seen that the azimuthal range next to the planet has a slower flow
than the azimuthal range opposite to the planet. Obeying the
continuity equation, the slower gas has to get denser, and the faster
gas thins. This situation is consistent with the density plots. 

To understand how a situation like \fig{fig:streamlines} can arise, we
perform a test placing massless particles at conjunction and at
opposition with a 5$M_J$ planet; the results are shown in
\fig{fig:eccentricities}. The star is at the center (orange dot in the
left panel) and the planet at $r=1$ (black dot in the left panel). We
place the particles initially at $r$=1.6, close but not exactly at the 2:1
resonance. The left panel shows two orbits of integration. The first
particle begins initially at opposition (at $\phi=0$, blue dot). As
the run starts, {its orbit becomes eccentric; at closest approach to
the planet it reaches aphelion (blue line). A circular orbit is shown as a dashed green
line for comparison. The second particle begins at the same radial
position but initially at conjunction (at $\phi=\pi$, red dot). Its
orbital eccentricity also grows, and at closest approach to the planet it also reaches aphelion, but at a slightly larger semimajor
axis (red line). Eccentricity and the semi-major axis do not
remain constant (middle and right panel of
\fig{fig:streamlines}), starting at eccentricity 
$e<$0.05 and later growing to as much as $e = 0.1$ for the outer and
$e =0.2$ for the inner particle. 
There is orbital variation, especially in the
case of the inner particle, due to the strong perturbation of the
planet. This precession will change on relatively longer timescales the argument of aphelion, and with
that the azimuthal location of the intensity maximum. We caution also
that the pressure field may change this picture.

Since the encounter with the planet occurs at aphelion and in this
region the motion is slower, we can expect the density there to be
higher. Taking the particle trajectories as a proxy for streamlines,
it seems that this eccentricity, caused by the massive planet deforming
the streamlines, can also lead to a lopsided azimuthal
overdensity. Although the eccentricity in our simulation is 
small, the saturation eccentricity of 0.25 predicted by \cite{KleyDirksen06} is far from negligible. The
hydrodynamic model will need to be run for far more than
the current 40 orbits in order for the eccentricity to saturate and
more accurate observational predictions of the outer gap edge to be
made. 


\subsection{Image catalog} 

We finish this work by presenting in the \ed{A}ppendix a catalog of images generated from our
simulations with shock heating, that may serve as a guide for
interpreting future observations. 

\section{Conclusions}
\label{sect:conclusions}

We have for the first time derived synthetic images 
\mm{including the heating produced by} the shocks that high-mass planets cause in
their parent disks. 
\mm{During this work,} the public
version of the code RADMC-3D was modified to include 
\mm{a general heating rate leading to emission from every cell.}

We explore the observational signatures of this shock heating, using
the hydrodynamical model of \citetalias{Lyra+16} of an
embedded 5$M_J$ planet, where the planetary shocks were first
characterized in three dimensions. We perform post-processing radiative transfer
simulations on this hydrodynamic model, 
finding that planetary shocks are able to match the general morphology
of the thermal emission in the
observations of HD 100546 from \cite{Currie+14}. Yet, synthetic 
images in the $L^\prime$ band generated by this post-processing
technique were able to 
\mm{match the observed morphology} only if 
an {\it ad hoc} factor 20 increase to the
shock heating rate is considered. 
At longer wavelengths (10\,$\mu$m), no {\it ad hoc} increase is
needed as the observational signature is prominent. \mm{We
  predict that at wavelengths of 10~$\mu$m or longer, the emission
  will come to be dominated by unpolarized thermal emission rather
  than polarized scattered emission.} 

We find that the best match to the observed feature \mm{in the $L'$
  band} comes not from
thermal radiation from the shock, but from scattered starlight off the
outer edge of the planetary gap. The gas immediately outward of the
planetary gap shows a raised scale height, with the surface of
$\tau=1$  matching primarily density, not temperature, contours, and lifted above the
average height of the disk atmosphere. Moreover, the density
enhancement is lopsided, with the side next to the planet being denser
and more prominent in the synthetic image than the side opposite to
the planet. The fact that scattering provides such a
good match to the observed features can be taken to imply that the
emission in HD 100456 as seen in \cite{Currie+14} is not thermal. 


In tracing the origin of the azimuthal asymmetry in the disk intensity  
outside the gap, we see that in the midplane the side next to the planet is indeed
denser and hotter than the opposite side. The maximum density in the
$\tau=1$ surface coincides with what looks at first as the eye of a
vortex. Yet, inspecting
other snapshots, the feature's location does not always match the
intensity maximum; moreover, the feature is stationary
in the reference frame of the planet, which led us to identify it with
a secondary spiral arm, excited at the 2:1 resonance. The cause of the asymmetry is ultimately the presence of this secondary
spiral arm intersecting with the disk gap outer edge. It increases the
density and temperature in the region, thus 
\mm{raising the height of the $\tau = 1$ surface there.}


If the outer gap edge is so prominent in thermodynamically evolving
disks with high-mass planets, we expect that eccentricity, if present, 
should have a significant observational signature. 
Measuring the streamlines we find that
the gas is faster in the opposite side and slower next to the planet,
as would be expected if the orbits were eccentric, with aphelion on
the planet side. Placing two test particles in orbit, initially in
circular Keplerian orbits, one starting in 
opposition and the other exterior conjunction, we see that, perturbed
by the planet, both quickly develop orbits with different semimajor
axes, but closest approach to
the planet in aphelion. Due to the strong
perturbation of the planet, the orbits evolve in time, and we do not
always expect the planet to be aligned with the orbital apsides. The
eccentricity is in fact just starting to develop after 40 orbits. \ed{It
will require simulating} the system for longer times to determine the
effect of a saturated eccentricity, which we
defer to a future study.

Future research to confirm the conclusions presented should vary the
distance to the planet in the hydrodynamic model, run the 
radiative transfer around a different type of star to match specific observations, and change the smoothing of planet mass. 
These studies would be individual to certain observations of spiral
emissions and should be tailored to each system. 
\mm{Measurements of polarization at wavelengths longer than 10~$\mu$m
  will reveal if thermal emission from the shock can indeed be
  detected.}

\acknowledgments{
B.H. acknowledges partial support 
from the Dobbs Ferry High School Science Research Program. W. L. acknowledges support of Space
Telescope Science Institute through grant HST-AR-14572 and the NASA
Exoplanet Research Program through grant 16-XRP16\_2-0065. The
simulations and post-processing presented in this paper utilized the Stampede
cluster of the Texas Advanced Computing Center (TACC) at The
University of Texas at Austin, and the Comet cluster at the University
of California at San Diego, both through XSEDE grant TG-AST140014.
This work was performed in part at the Jet Propulsion
Laboratory. We acknowledge informative discussions with Konstantin Batygin, 
Aaron Boley, Thayne Currie, Robin Dong, Cornelis Dullemond, Max
Millar-Blanchaer, and Zsolt S\'andor.}

\bibliographystyle{apj}


\appendix

We present here a catalog of images generated from our
simulations with shock heating, that may serve as a guide to future observations. 

In \fig{fig:3,5scat-array} we show images calculated at 3.5\,$\mu$m,
with scattering included. From top to bottom the position angles are 0\degree, 90\degree,
180\degree, and 270\degree. From left to right the images are shown in inclination angles of 0\degree, 45\degree, and 80\degree.

In \fig{fig:10scat-array} we show images in the same orientations as
\fig{fig:3,5scat-array}, but at 10\,$\mu$m and with the original
heating rate. In Figs \ref{fig:3,5shock-array} and \ref{fig:10shock-array} we show the
same images but with scattering excluded, i.e., only the effect of
shocks. 

Comparison between \fig{fig:3,5scat-array} and
\fig{fig:3,5shock-array} shows that scattering is by far the main
feature in this waveband. The luminosity from the shocks, faint and
deep within the midplane, is orders of magnitude smaller than the
scattering off the disk surface. 

\begin{figure*}
  \begin{center}
    \resizebox{.25\textwidth}{!}{\includegraphics{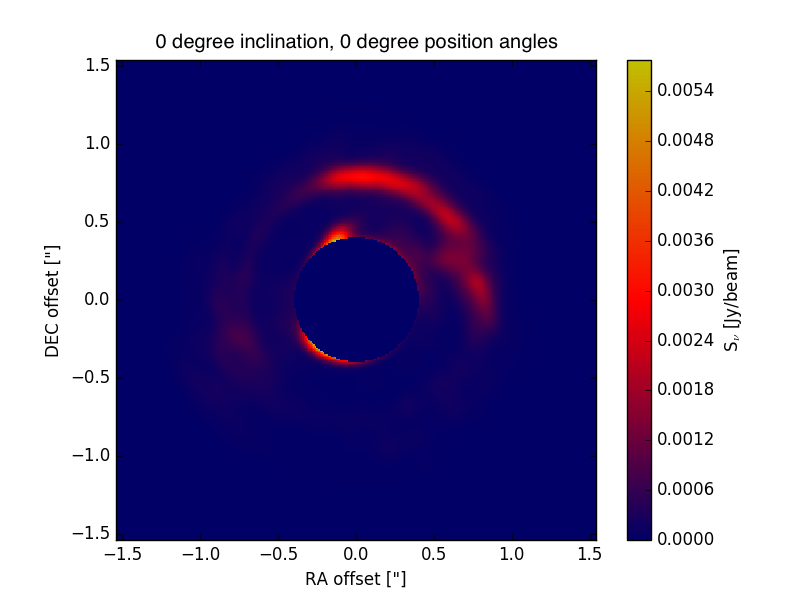}}
    \resizebox{.25\textwidth}{!}{\includegraphics{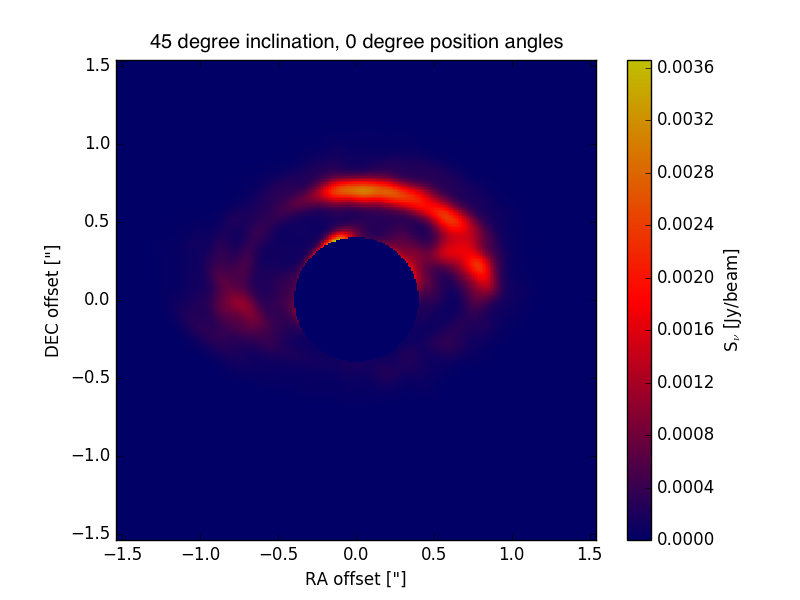}}
    \resizebox{.25\textwidth}{!}{\includegraphics{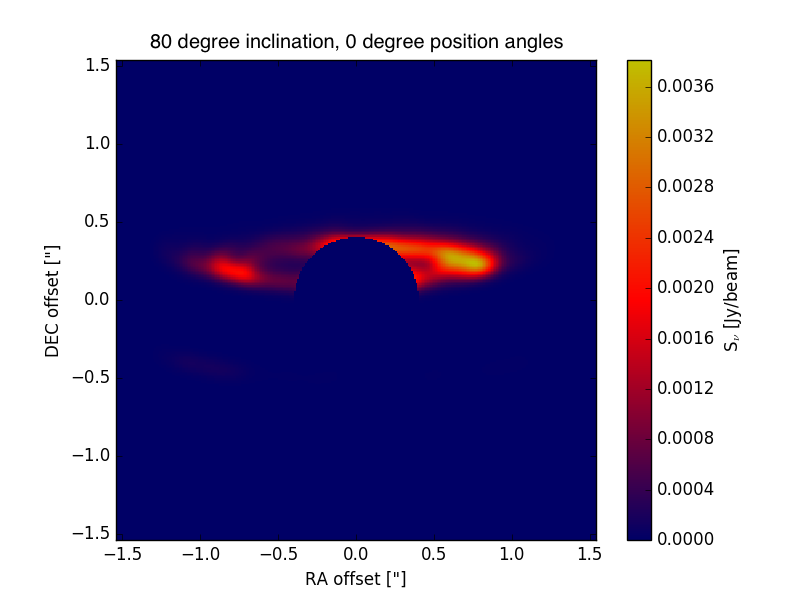}}

    \resizebox{.25\textwidth}{!}{\includegraphics{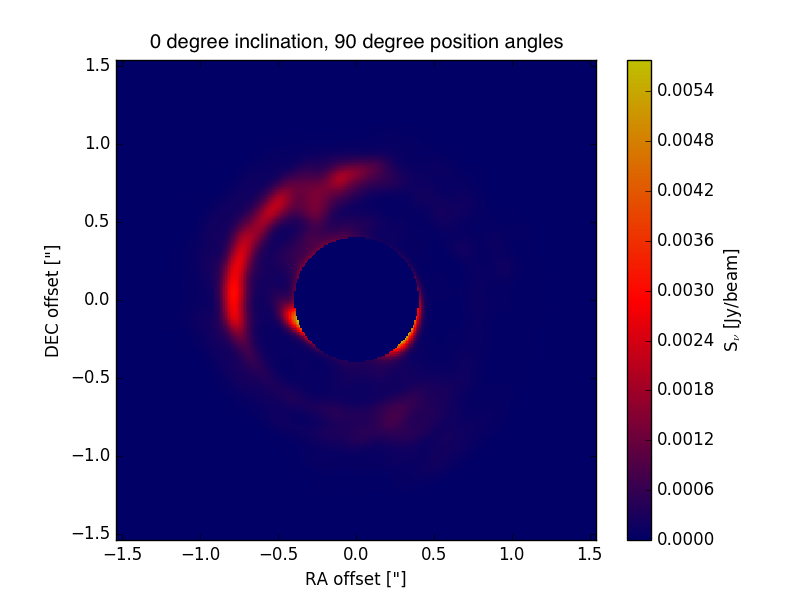}}
    \resizebox{.25\textwidth}{!}{\includegraphics{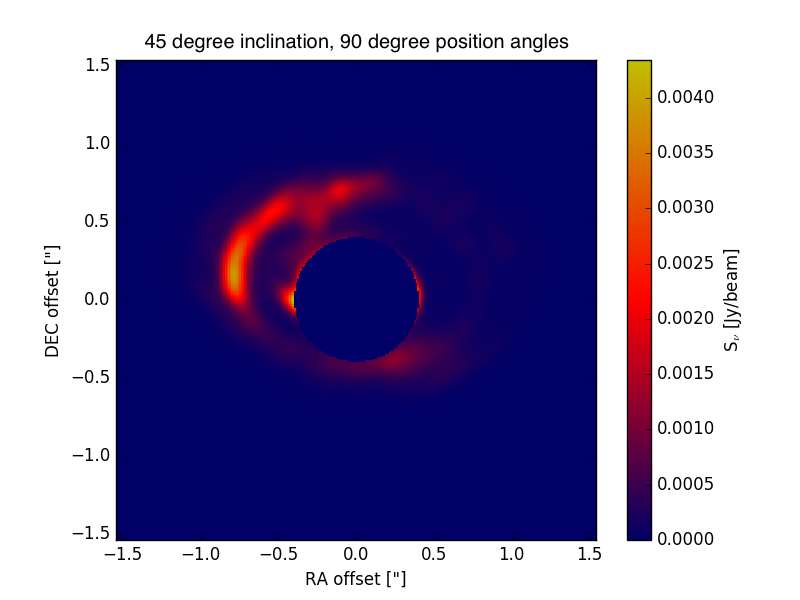}}
    \resizebox{.25\textwidth}{!}{\includegraphics{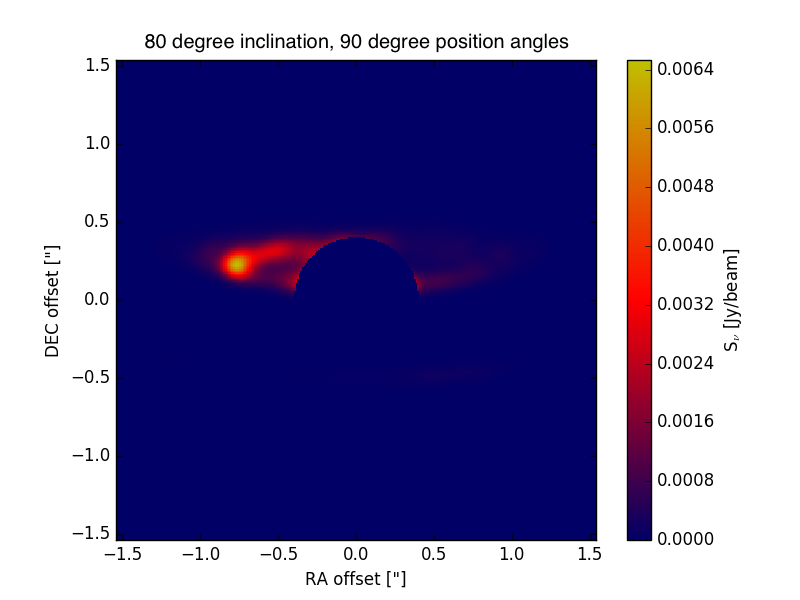}}

    \resizebox{.25\textwidth}{!}{\includegraphics{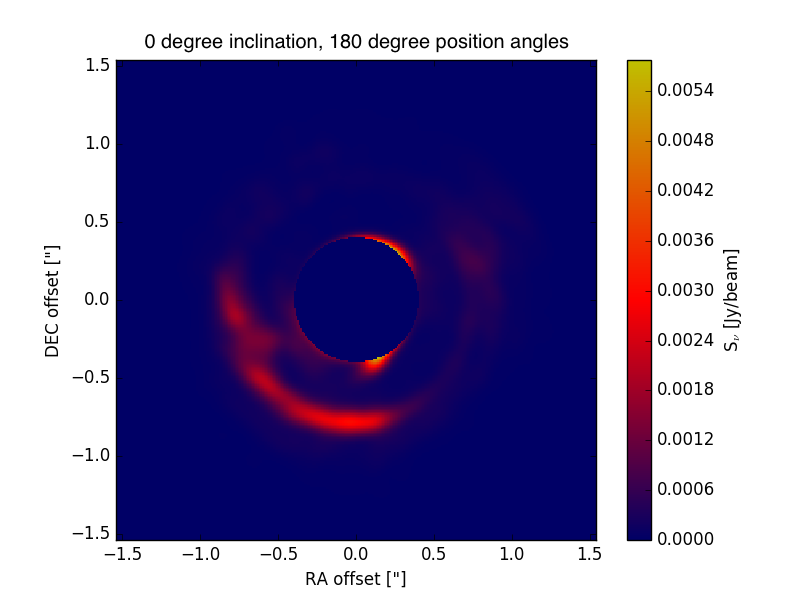}}
    \resizebox{.25\textwidth}{!}{\includegraphics{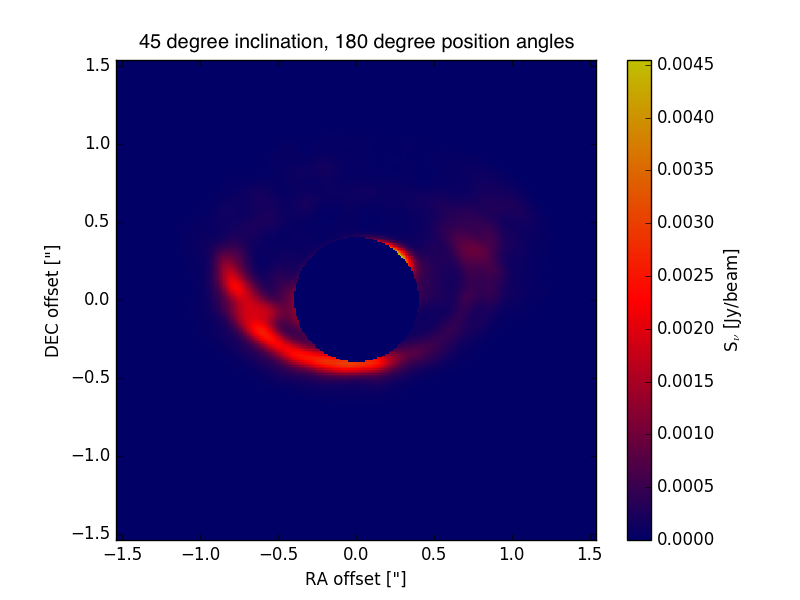}}
    \resizebox{.25\textwidth}{!}{\includegraphics{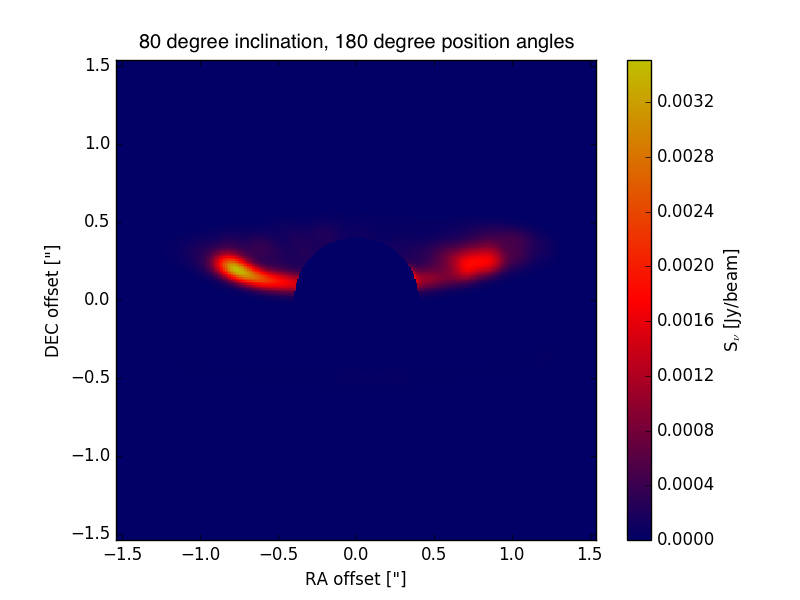}}

    \resizebox{.25\textwidth}{!}{\includegraphics{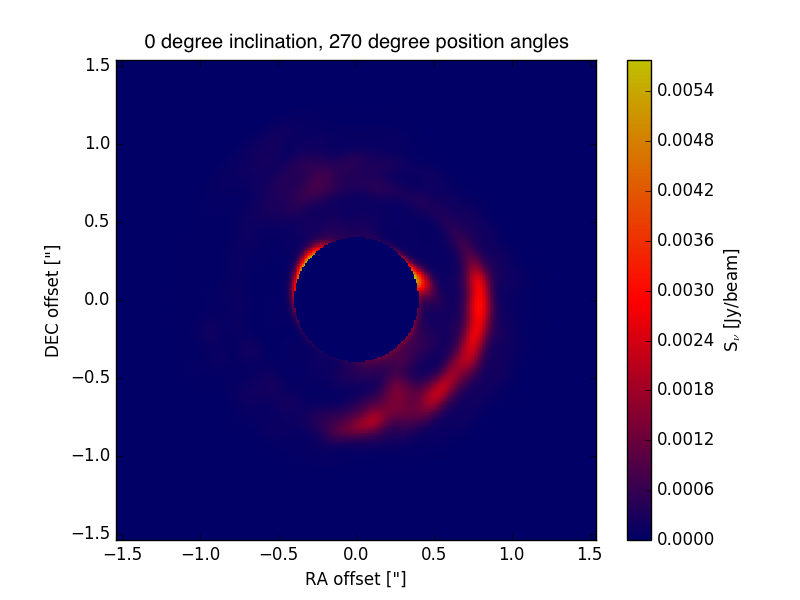}}
    \resizebox{.25\textwidth}{!}{\includegraphics{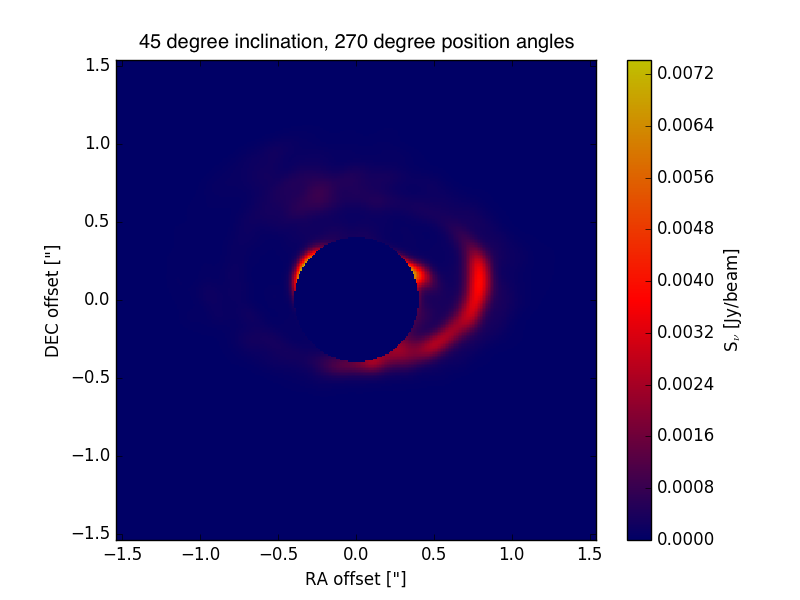}}
    \resizebox{.25\textwidth}{!}{\includegraphics{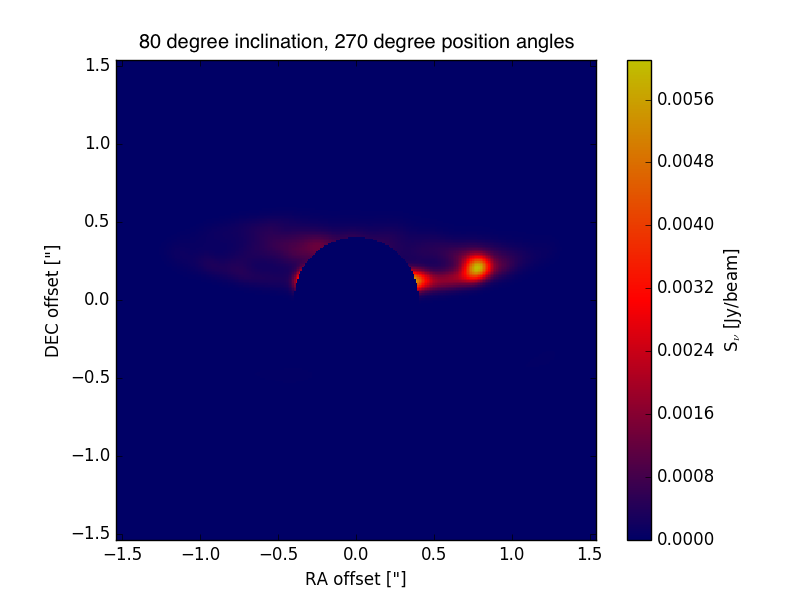}}
\end{center}
\caption[]{Array of synthetic images at 3.5~$\mu$m with scattering
  included. The \ed{original shock heating rate is used}. Rows
  from top to bottom are position angles of 0\degree, 90\degree,
  180\degree, and 270\degree. From left to right are inclination
  angles of 0\degree, 45\degree, and 80\degree.}
\label{fig:3,5scat-array}
\end{figure*}

\begin{figure*}
  \begin{center}
    \resizebox{.25\textwidth}{!}{\includegraphics{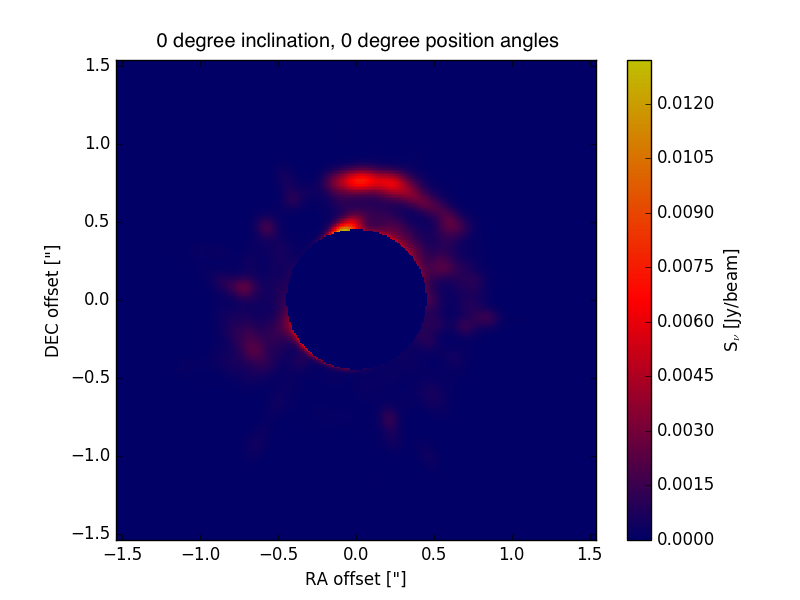}}
    \resizebox{.25\textwidth}{!}{\includegraphics{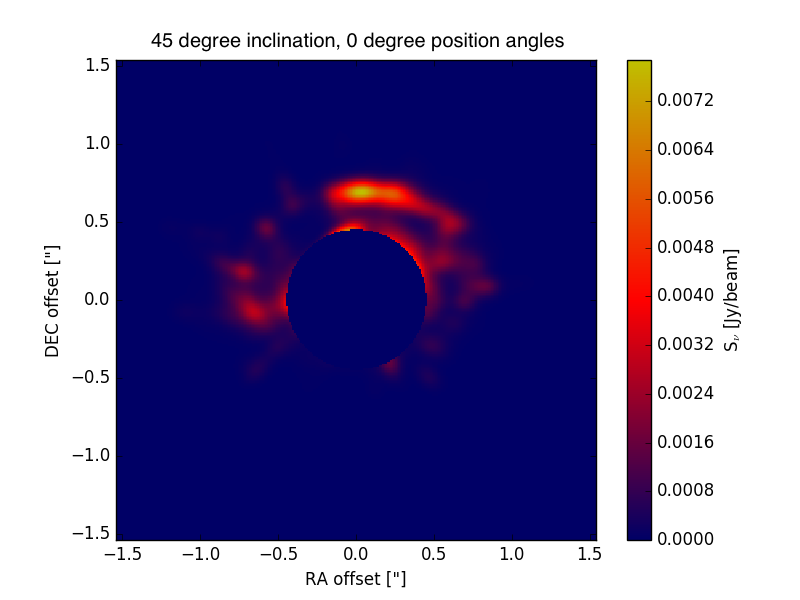}}
    \resizebox{.25\textwidth}{!}{\includegraphics{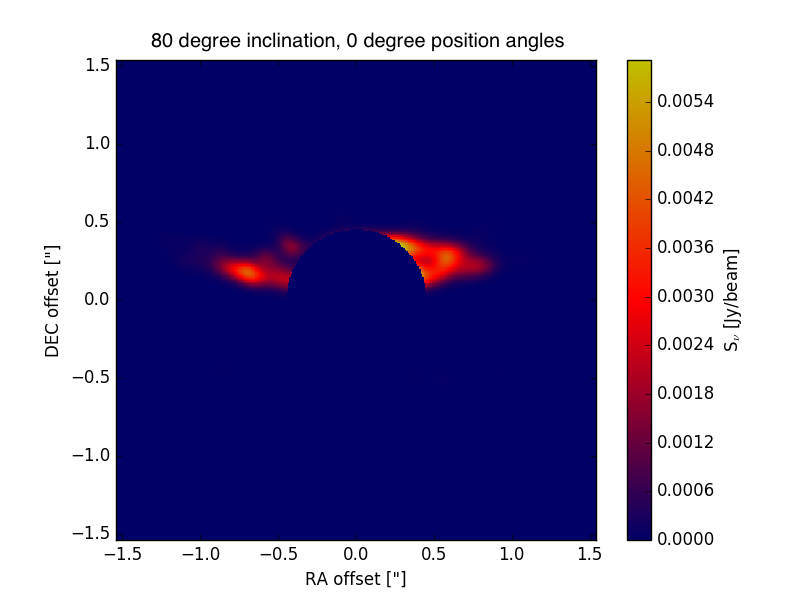}}

    \resizebox{.25\textwidth}{!}{\includegraphics{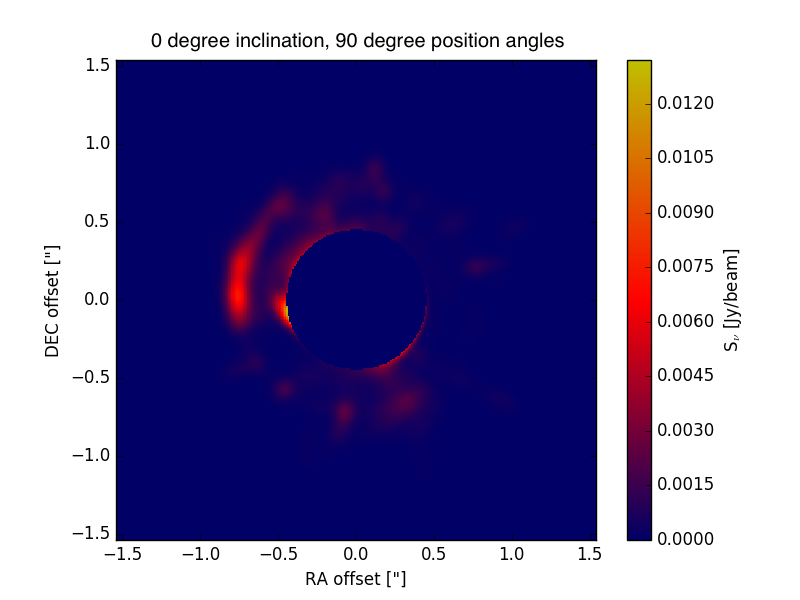}}
    \resizebox{.25\textwidth}{!}{\includegraphics{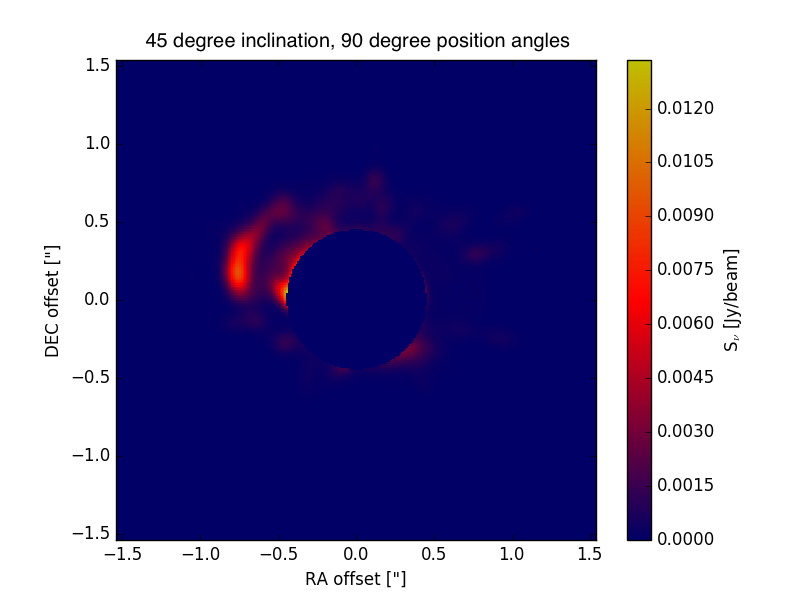}}
    \resizebox{.25\textwidth}{!}{\includegraphics{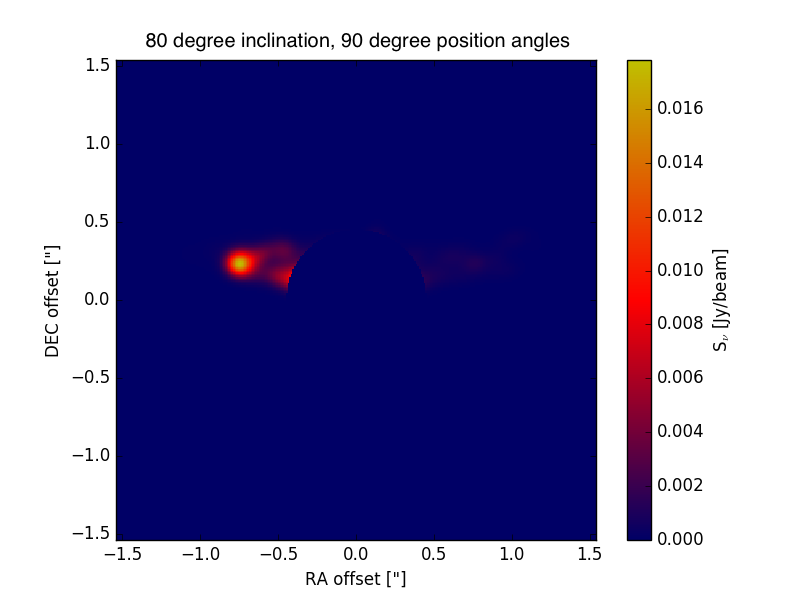}}

    \resizebox{.25\textwidth}{!}{\includegraphics{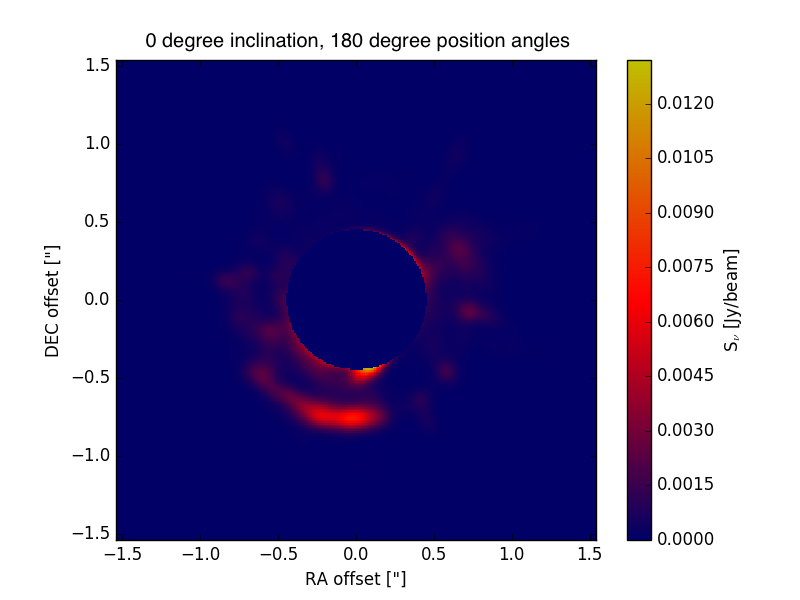}}
    \resizebox{.25\textwidth}{!}{\includegraphics{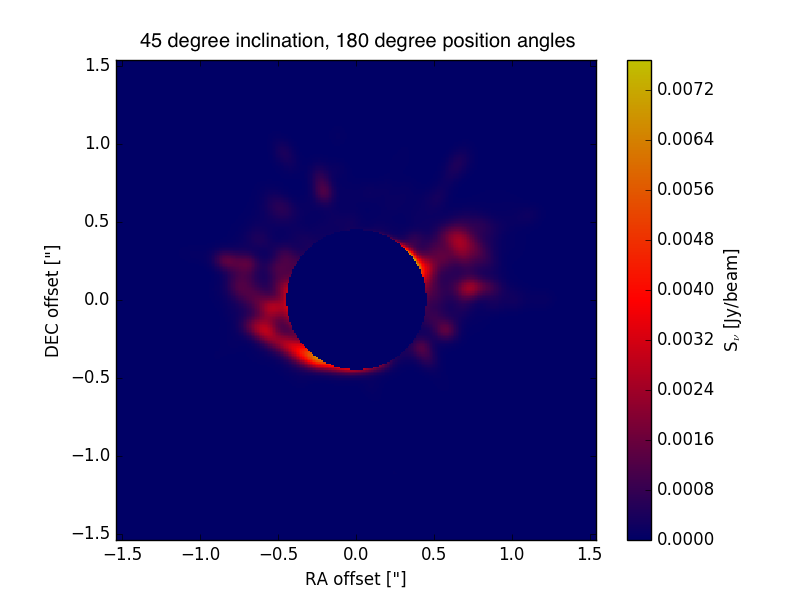}}
    \resizebox{.25\textwidth}{!}{\includegraphics{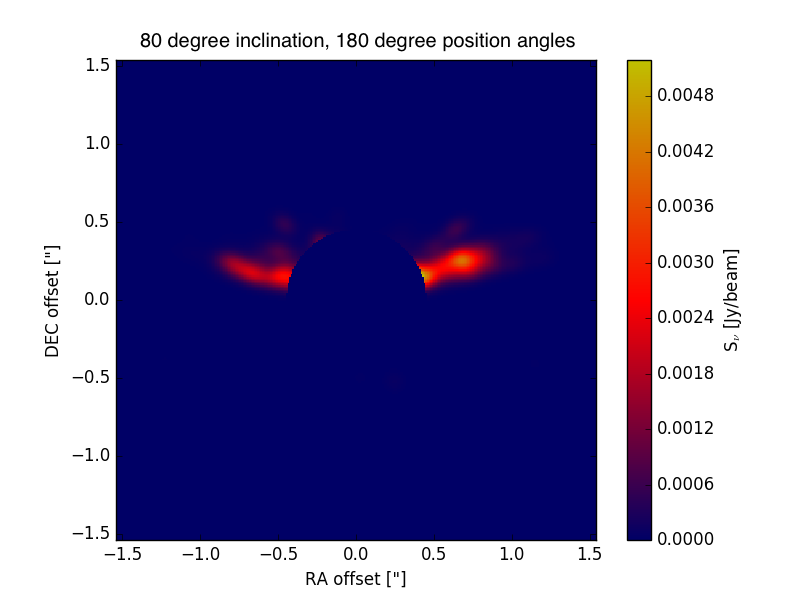}}

    \resizebox{.25\textwidth}{!}{\includegraphics{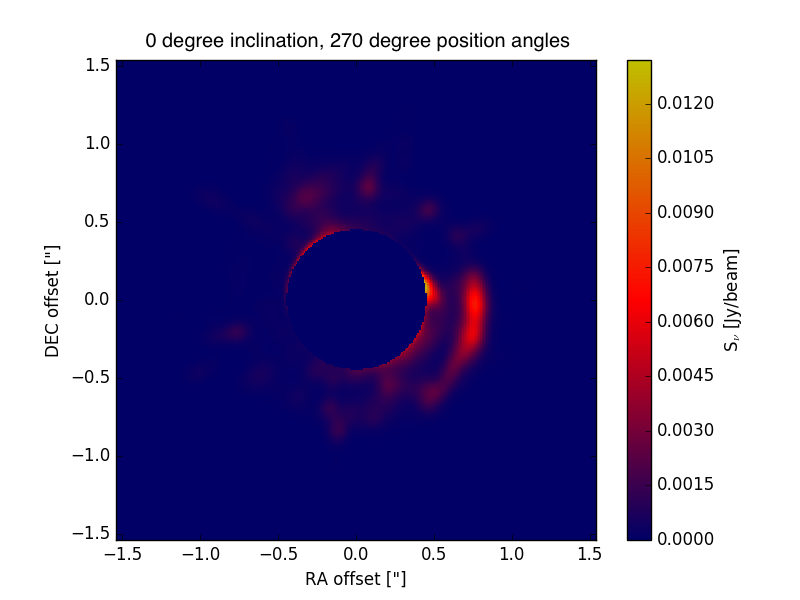}}
    \resizebox{.25\textwidth}{!}{\includegraphics{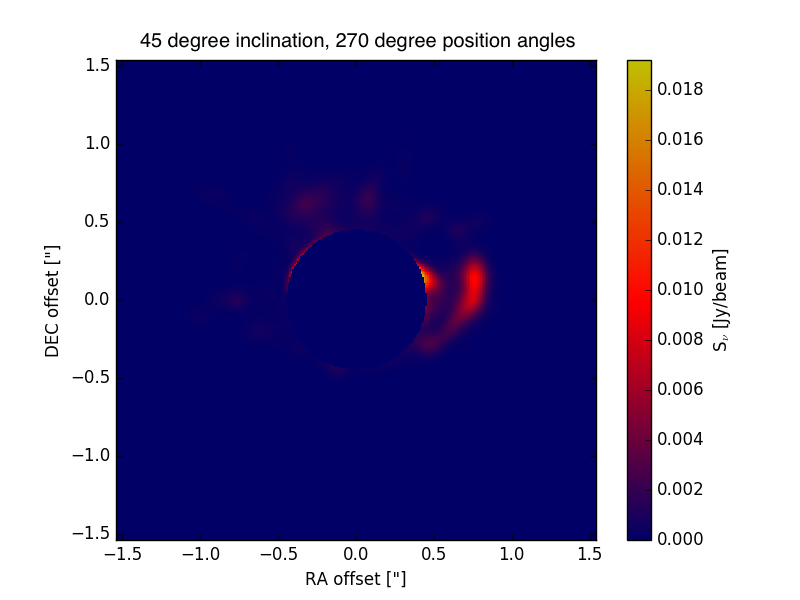}}
    \resizebox{.25\textwidth}{!}{\includegraphics{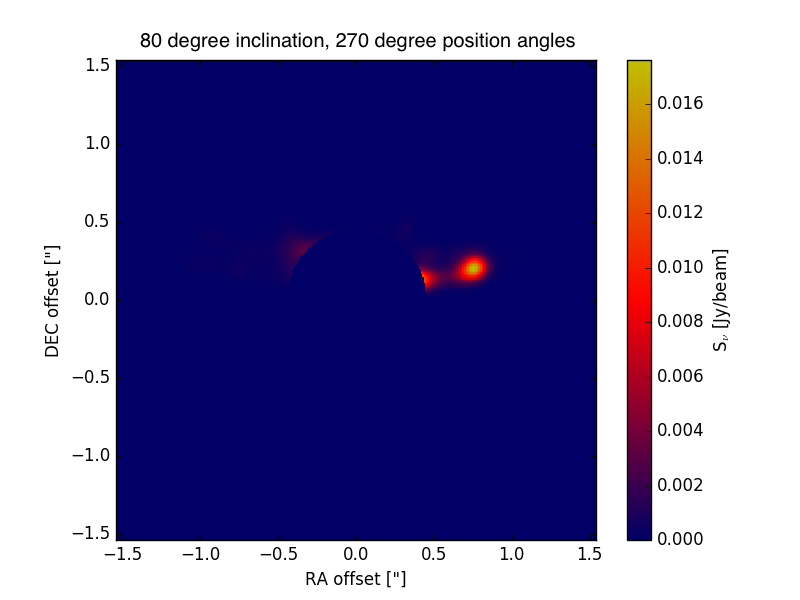}}
\end{center}
\caption[]{Array of synthetic images at 10~$\mu$m with scattering
  included. The original shock heating rate is used. Rows
  from top to bottom are position angles of 0\degree, 90\degree,
  180\degree, and 270\degree. From left to right are inclination
  angles of 0\degree, 45\degree, and 80\degree.}
\label{fig:10scat-array}
\end{figure*} 

\begin{figure*}
  \begin{center}
    \resizebox{.25\textwidth}{!}{\includegraphics{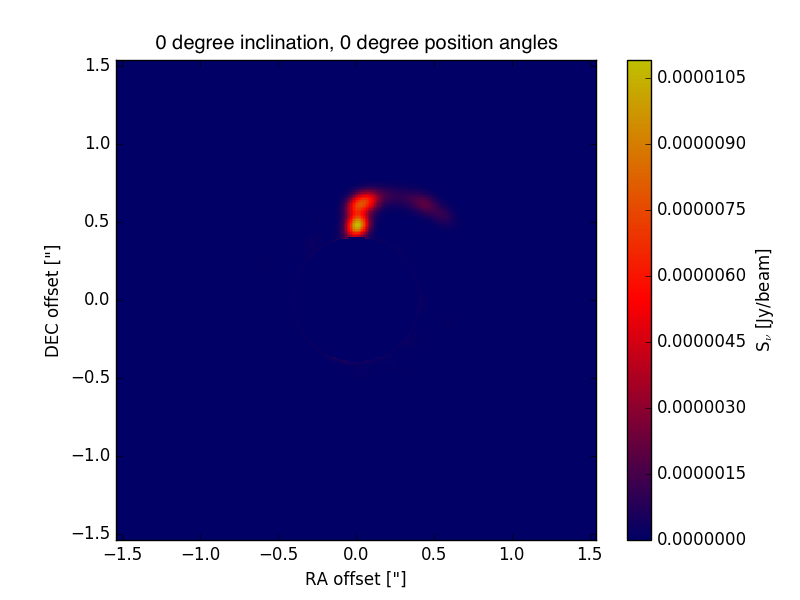}}
    \resizebox{.25\textwidth}{!}{\includegraphics{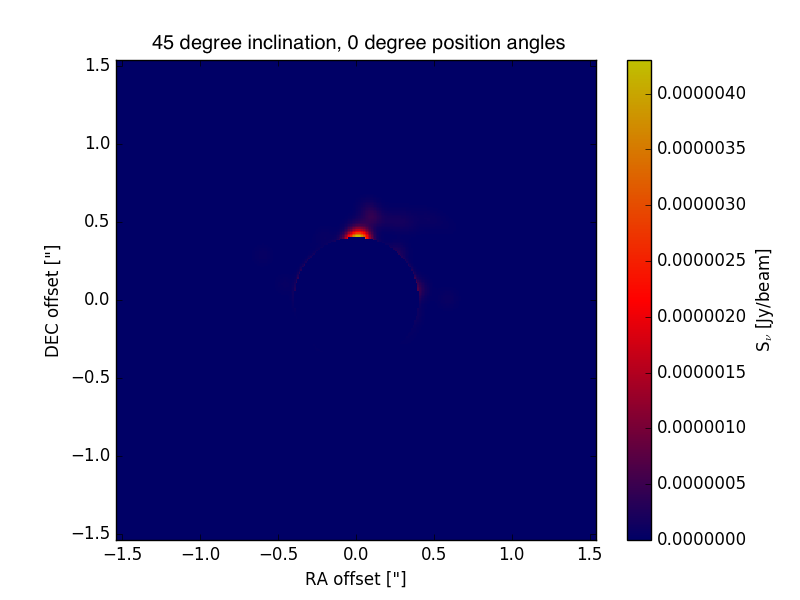}}
    \resizebox{.25\textwidth}{!}{\includegraphics{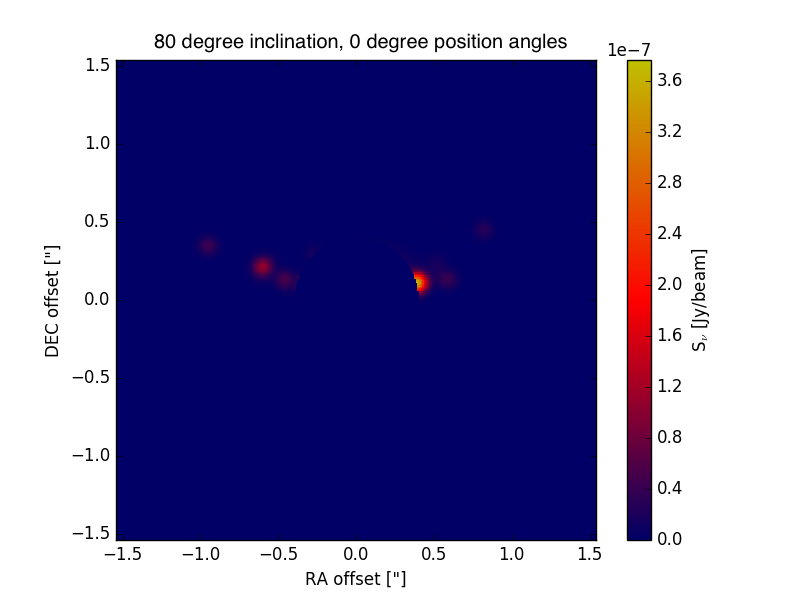}}

    \resizebox{.25\textwidth}{!}{\includegraphics{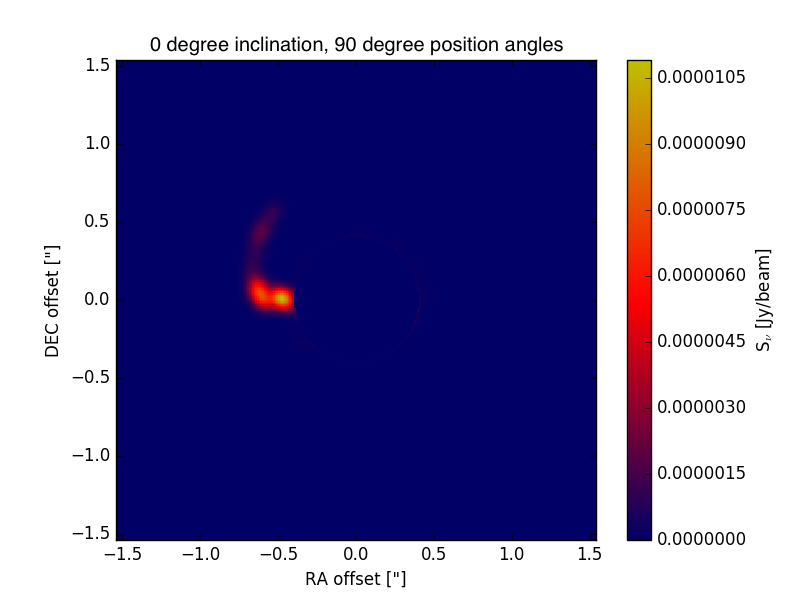}}
    \resizebox{.25\textwidth}{!}{\includegraphics{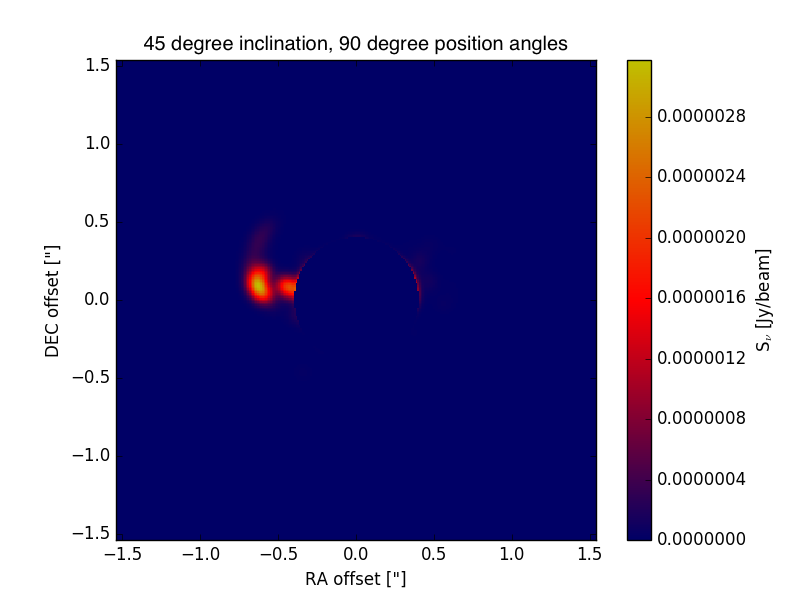}}
    \resizebox{.25\textwidth}{!}{\includegraphics{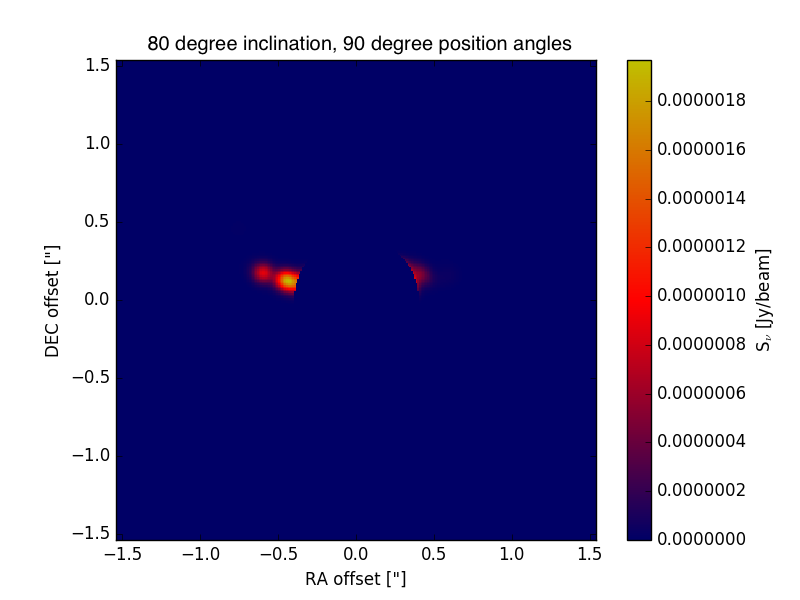}}

    \resizebox{.25\textwidth}{!}{\includegraphics{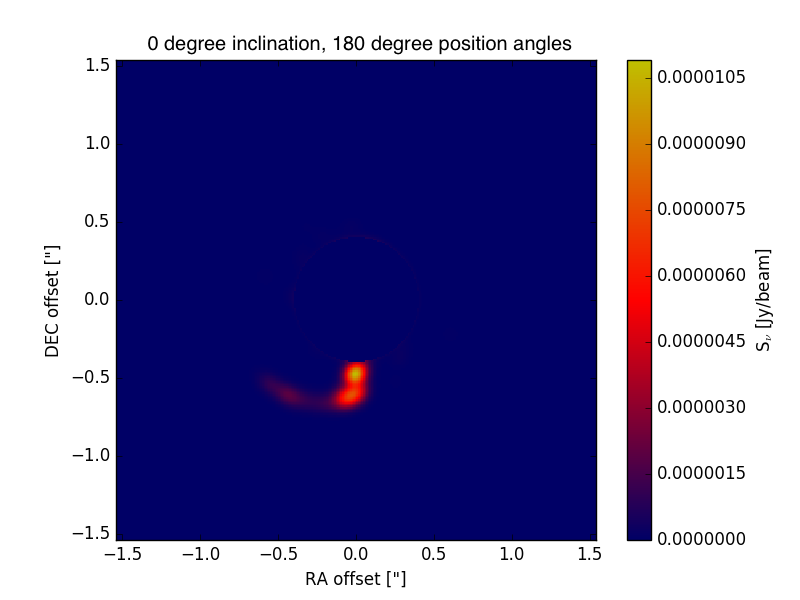}}
    \resizebox{.25\textwidth}{!}{\includegraphics{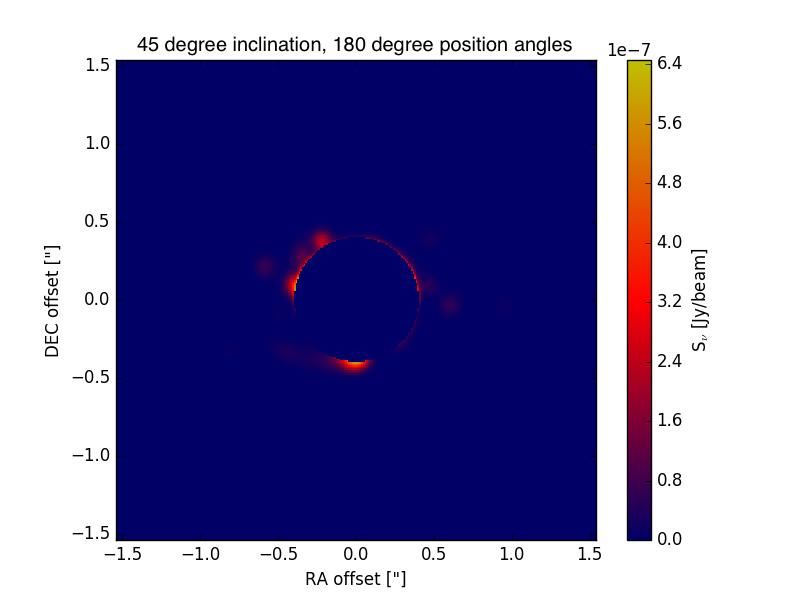}}
    \resizebox{.25\textwidth}{!}{\includegraphics{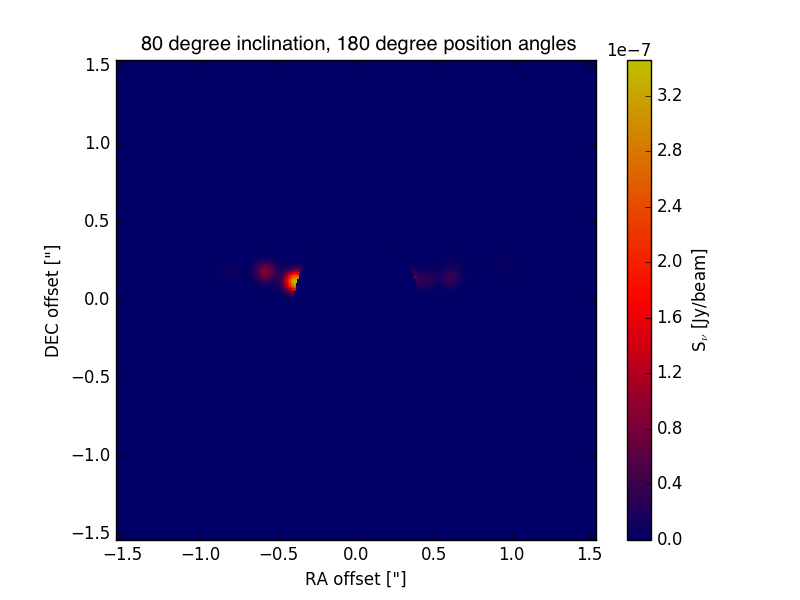}}

    \resizebox{.25\textwidth}{!}{\includegraphics{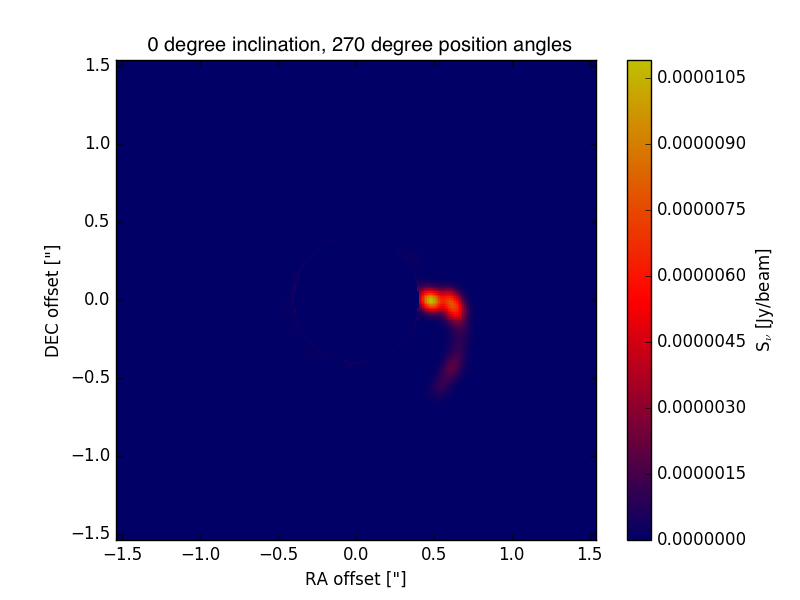}}
    \resizebox{.25\textwidth}{!}{\includegraphics{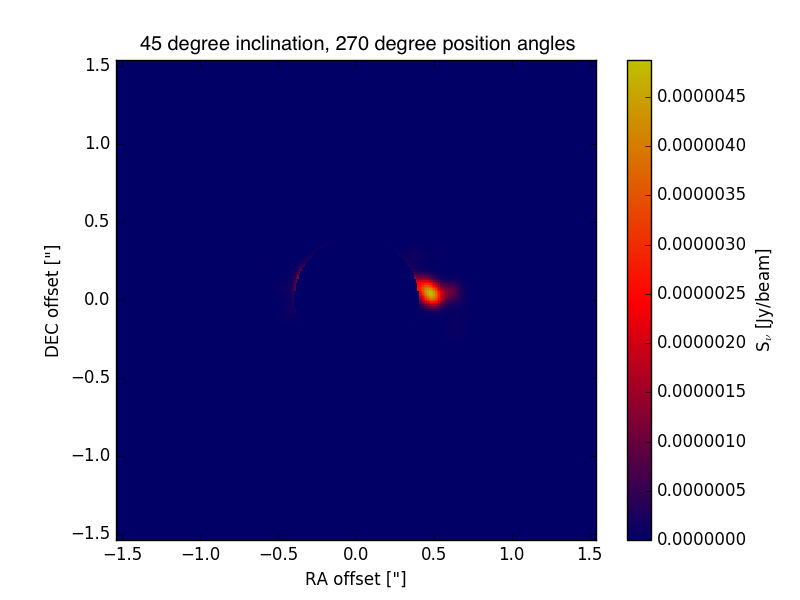}}
    \resizebox{.25\textwidth}{!}{\includegraphics{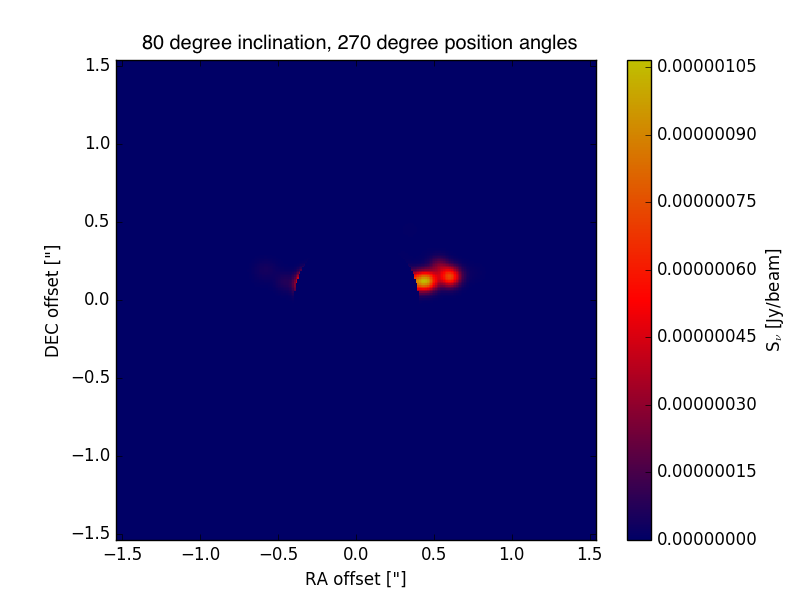}}
\end{center}
\caption[]{Array of synthetic images at 3.5~$\mu$m without scattering
  included.  The shock heating rate is increased by a factor 20,
  \ed{since images produced with the original shock heating rate
    show no distinguishable features}. Rows
  from top to bottom are position angles of 0\degree, 90\degree,
  180\degree, and 270\degree. From left to right are inclination
  angles of 0\degree, 45\degree, and 80\degree.}
\label{fig:3,5shock-array}
\end{figure*} 

\begin{figure*}
  \begin{center}
    \resizebox{.25\textwidth}{!}{\includegraphics{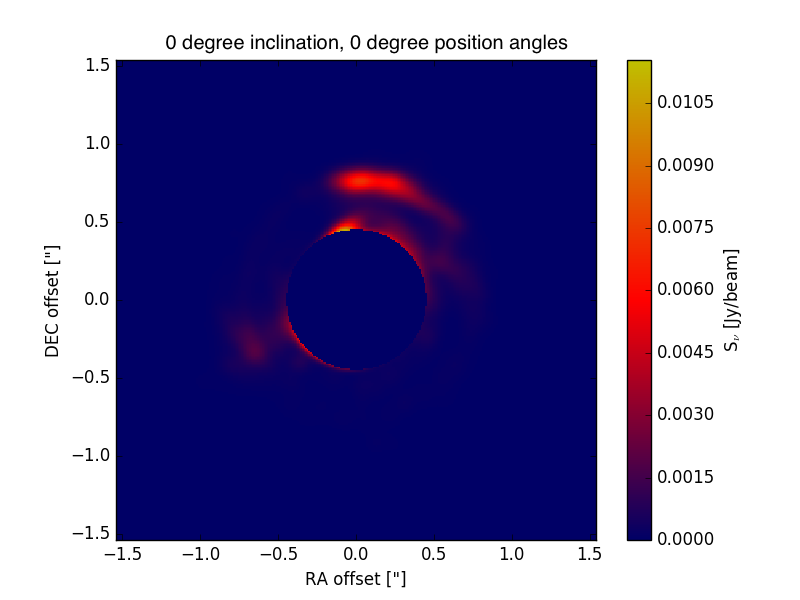}}
    \resizebox{.25\textwidth}{!}{\includegraphics{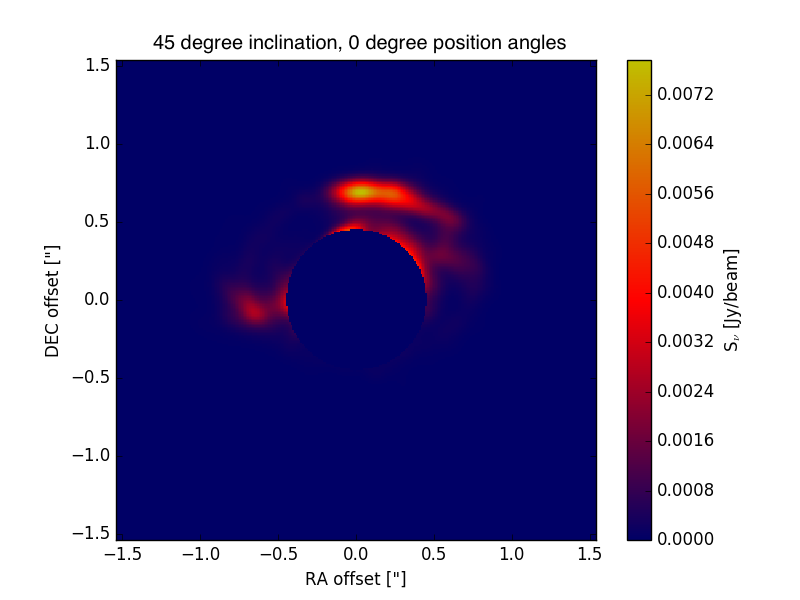}}
    \resizebox{.25\textwidth}{!}{\includegraphics{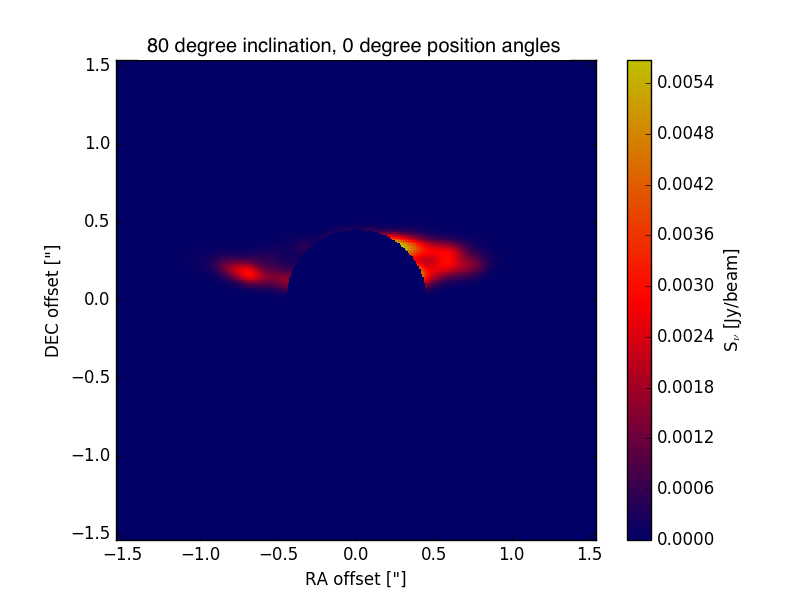}}

    \resizebox{.25\textwidth}{!}{\includegraphics{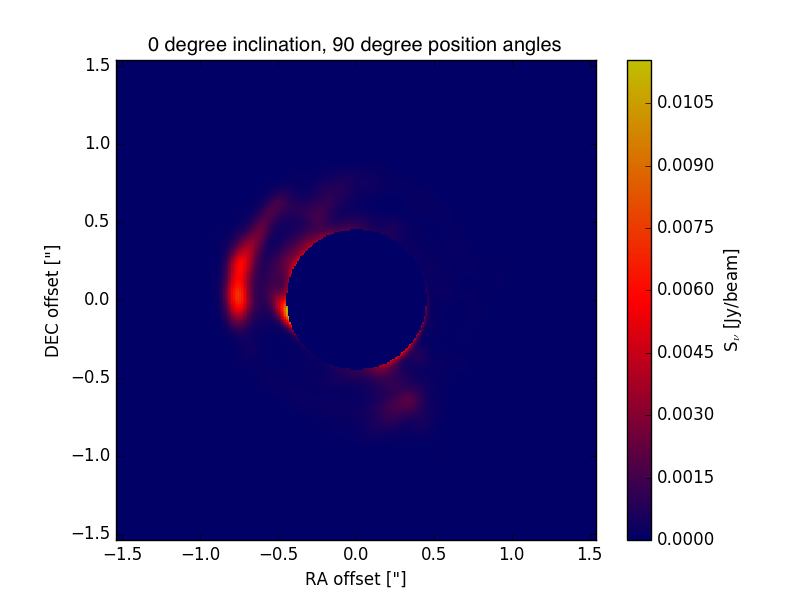}}
    \resizebox{.25\textwidth}{!}{\includegraphics{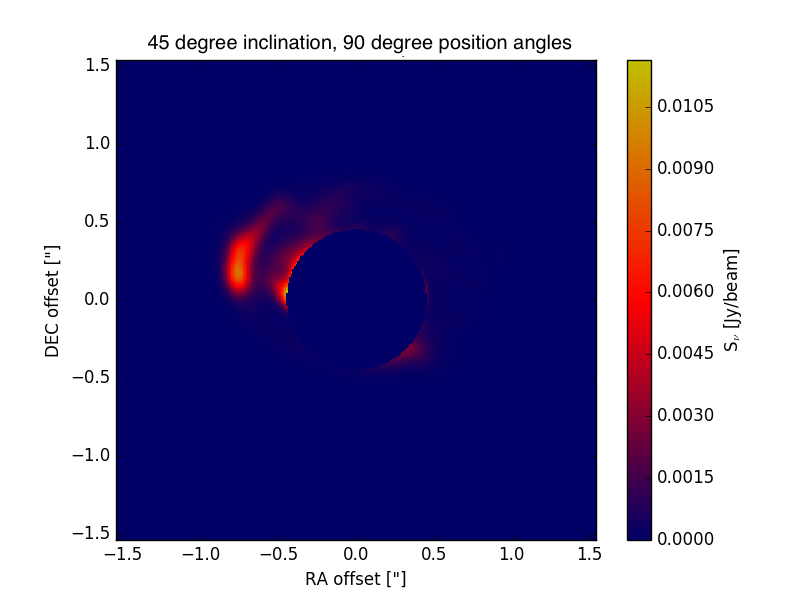}}
    \resizebox{.25\textwidth}{!}{\includegraphics{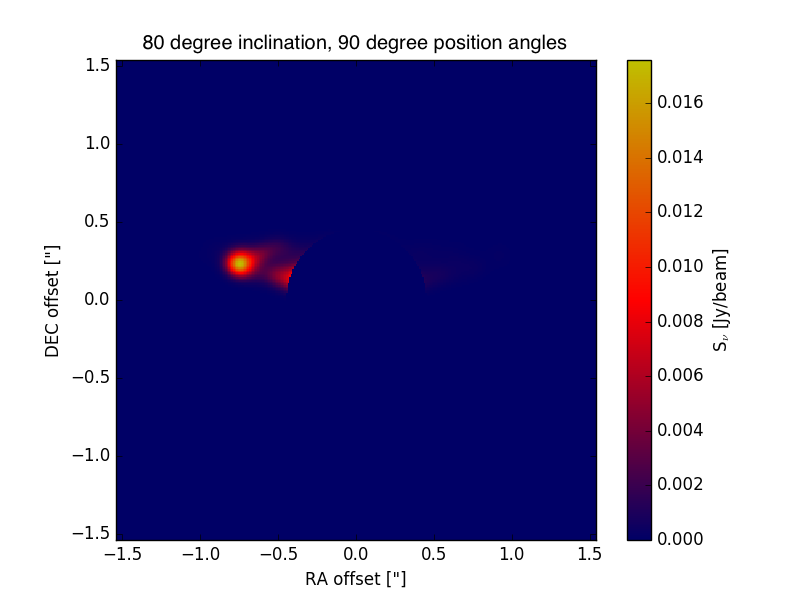}}

    \resizebox{.25\textwidth}{!}{\includegraphics{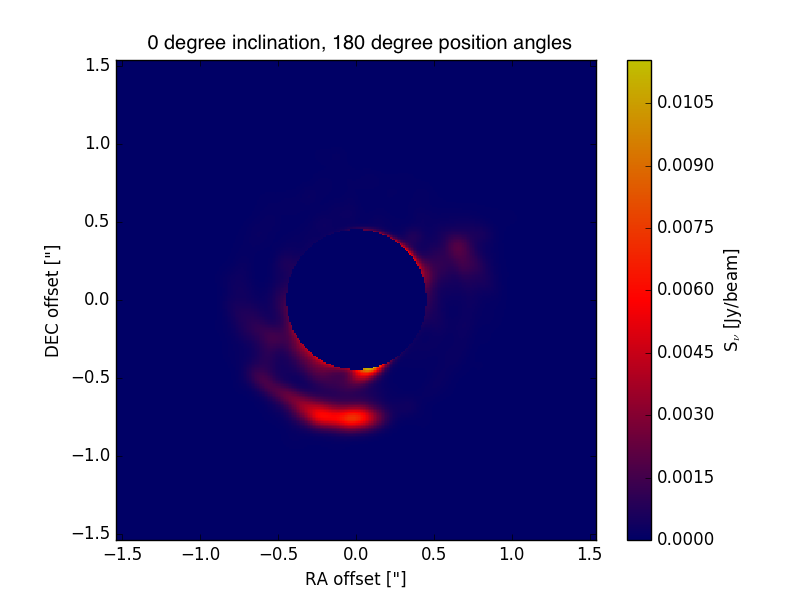}}
    \resizebox{.25\textwidth}{!}{\includegraphics{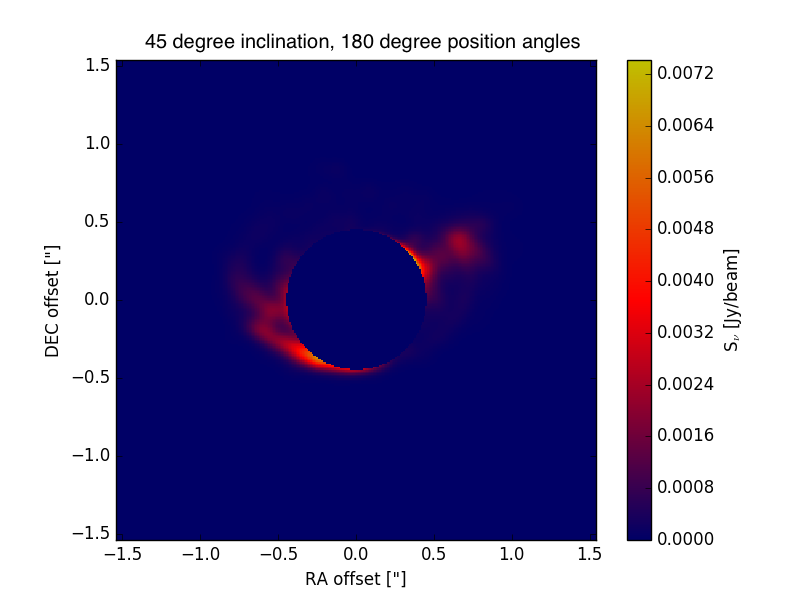}}
    \resizebox{.25\textwidth}{!}{\includegraphics{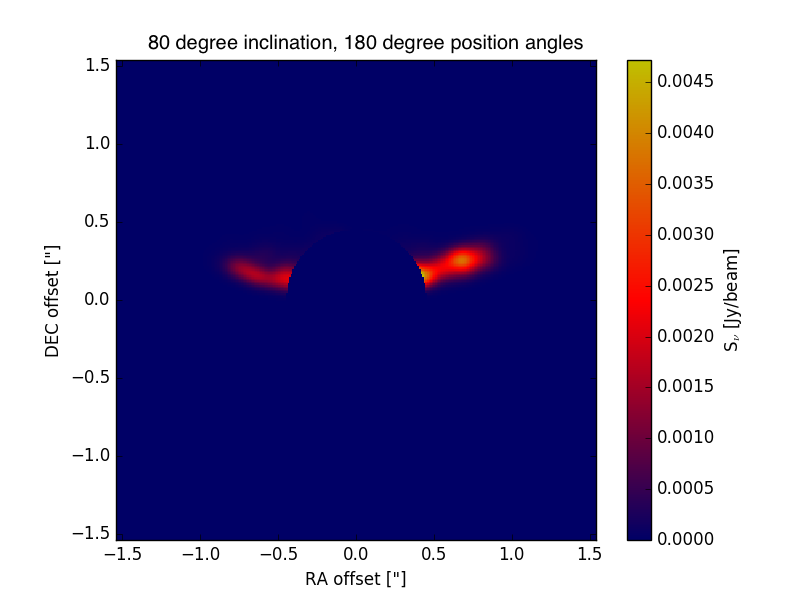}}

    \resizebox{.25\textwidth}{!}{\includegraphics{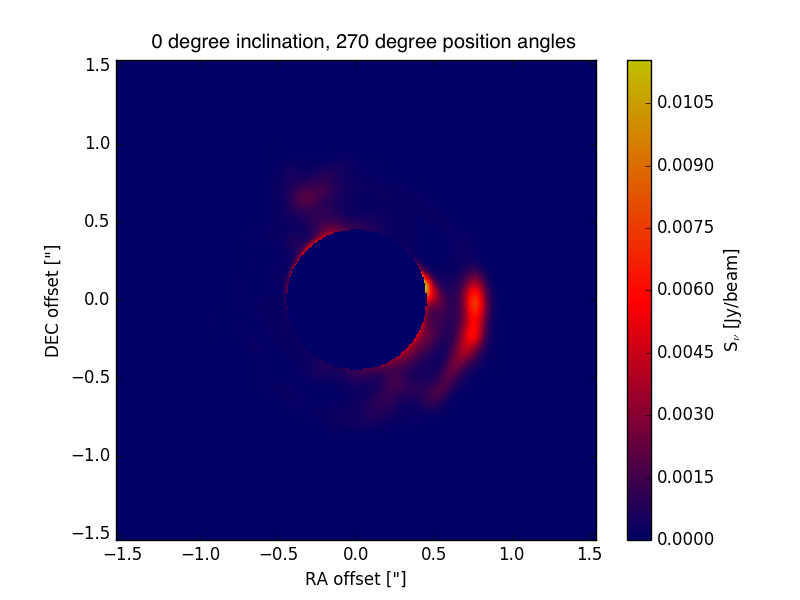}}
    \resizebox{.25\textwidth}{!}{\includegraphics{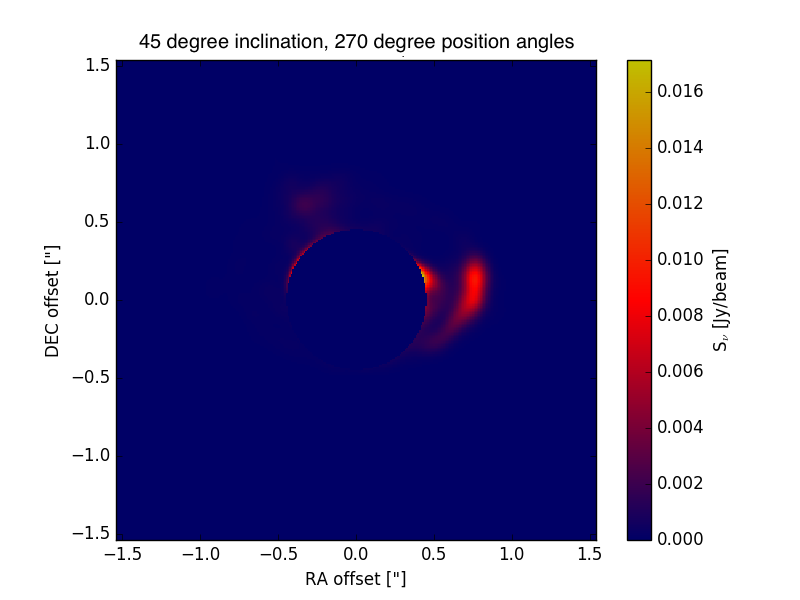}}
    \resizebox{.25\textwidth}{!}{\includegraphics{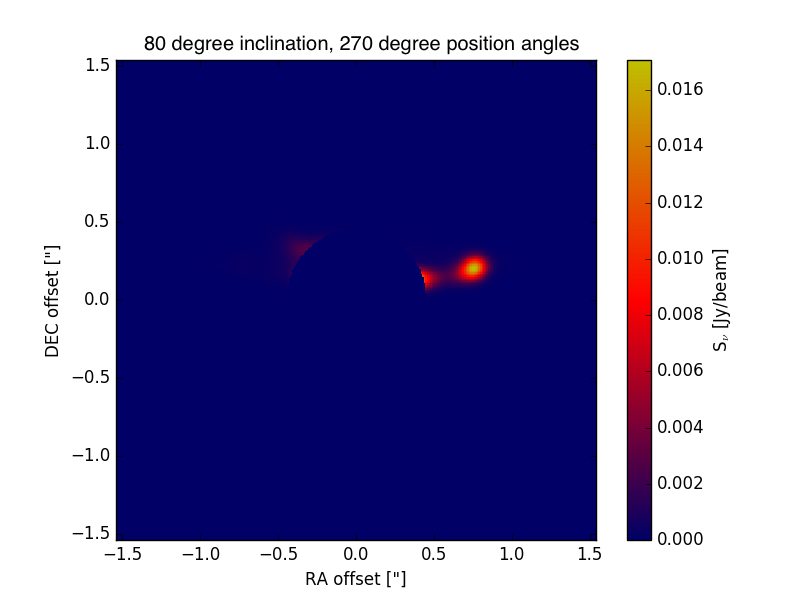}}
\end{center}
\caption[]{Array of synthetic images at 10~$\mu$m without scattering
  included. The original shock heating rate is used. Rows
  from top to bottom are position angles of 0\degree, 90\degree,
  180\degree, and 270\degree. From left to right are inclination
  angles of 0\degree, 45\degree, and 80\degree.}
\label{fig:10shock-array}
\end{figure*}

\end{document}